\pdfoutput=1

\documentclass[prb,superscriptaddress,amsmath,amssymb,twocolumn]{revtex4-1}

\usepackage{graphicx,bm,amsmath}
\usepackage[usenames,dvipsnames]{xcolor}
\usepackage[colorlinks,bookmarks=false,citecolor=NavyBlue,linkcolor=Red,urlcolor=blue]{hyperref}
\usepackage[bbgreekl]{mathbbol}
\usepackage{xcolor}
\usepackage[normalem]{ulem}
\usepackage{braket}
\usepackage{bm}
\usepackage{esint}

\def\doi{http://dx.doi.org/}

\newcommand{\be}{\begin{equation}}
\newcommand{\ee}{\end{equation}}
\newcommand{\bea}{\begin{eqnarray}}
\newcommand{\eea}{\end{eqnarray}}

\def\dd{{\rm d}}

\newcommand{\ocite}[1]{[\onlinecite{#1}]}

\newcommand{\titleinfo}{Adiabatic formation of bound states in the 1d Bose gas}

\begin{document}

\title{\titleinfo}

\author{Rebekka Koch}
\affiliation{Institute for Theoretical Physics, University of Amsterdam, PO Box 94485, 1090 GL Amsterdam, The Netherlands}

\author{Alvise Bastianello}
\affiliation{Institute for Theoretical Physics, University of Amsterdam, PO Box 94485, 1090 GL Amsterdam, The Netherlands}
\affiliation{Department of Physics and Institute for Advanced Study, Technical University of Munich, 85748 Garching, Germany}
\affiliation{Munich Center for Quantum Science and Technology (MCQST), Schellingstr. 4, D-80799 M{\"u}nchen, Germany}

\author{Jean-S\'ebastien Caux}
\affiliation{Institute for Theoretical Physics, University of Amsterdam, PO Box 94485, 1090 GL Amsterdam, The Netherlands}

\begin{abstract}
We consider the 1d interacting Bose gas in the presence of time-dependent and spatially inhomogeneous contact interactions. Within its attractive phase, the gas allows for bound states of an arbitrary number of particles, which are eventually populated if the system is dynamically driven from the repulsive to the attractive regime. Building on the framework of Generalized Hydrodynamics, we analytically determine the formation of bound states in the limit of adiabatic changes in the interactions.
Our results are valid for arbitrary initial thermal states and, more generally, Generalized  Gibbs Ensembles. 
\end{abstract}

\maketitle

\section{Introduction}

Many-body quantum systems are extremely sensitive to interactions, leading to a wide variety of possible phases of matter.  This is particularly evident in low-dimensional systems, where particles are forced to meet, therefore to scatter: hence, the tiniest modification in the interactions can lead to deep physical changes.
The 1d world is nowadays routinely probed in the lab, thanks to the astonishing advances in the context of cold atoms \ocite{Bloch2005}: several out-of-equilibrium protocols have been engineered, unveiling new phases of matter \ocite{Gring1318,Schneider2012,Langen207}. Local observables and correlations thereof are measured in great detail thanks to in-situ manipulations \ocite{Fukuhara2013,Jepsen2020,PhysRevLett.113.147205,Fukuhara2013_bis}.
As one of the main protagonists in this play the Bose gas with contact interactions, also known as the Lieb-Liniger model (LL) \ocite{PhysRev.130.1605,PhysRev.130.1616,doi:10.1063/1.1704156}, stands out since it naturally emerges when a bosonic gas is confined to an elongated trap \ocite{PhysRevLett.81.938,Bloch2005,RevModPhys.80.885,Kinoshita1125,PhysRevLett.95.190406,Kinoshita2006,PhysRevLett.105.230402,PhysRevLett.100.090402,PhysRevLett.122.090601,PhysRevLett.100.090402,PhysRevLett.107.230404,malvania2020generalized,Haller1224,PabloSolanoExp2019,kao2020creating}.
The LL model belongs to the class of integrable, or exactly solvable, systems \ocite{korepin1997quantum,Mussardo:2010mgq} which possess an extensive number of local conserved quantities: this has far-reaching consequences such as hindering of thermalization \ocite{RevModPhys.83.863,Calabrese_2016} and ballistic transport \ocite{bertini2020finitetemperature}. Indeed, homogeneous integrable models relax to a non-equilibrium steady state known as Generalized Gibbs Ensemble (GGE) \ocite{rigol2007relaxation}, which is sensitive to the whole set of dynamical constraints.

The interaction of the LL model can be experimentally tuned with great accuracy  through Feshbach resonances \ocite{Inouye1998} or through trap squeezing \ocite{PhysRevLett.81.938,PhysRevLett.91.163201}, allowing experimentalists to probe an interesting dichotomy in its phase space. Indeed, depending on the sign of the interaction, the LL's excitation content completely changes: in contrast to the repulsive phase, the attractive one sustains stable bound states of an arbitrary number of particles \ocite{takahashi2005thermodynamics,
PhysRevLett.98.150403,
Calabrese_2007,PhysRevA.72.033613}.
In the attractive phase, the homogeneous ground state is a single massive molecula with overextensive energy $\propto -N^3$, with $N$ the number of particles \ocite{doi:10.1063/1.1704156,PhysRevA.11.265}. This exotic state has been thoroughly investigated, with particular emphasis on its instability against local perturbations \ocite{PhysRevA.73.033611,CASTIN2001419} and effects of traps \ocite{Tempfli_2008}.
On the other hand, in out-of-equilibrium setups with extensive energy a well-defined thermodynamic limit in the usual sense is restored, sparking the interest on both the theoretical and experimental sides.
The dynamical production of bound states due to interaction changes is of primary experimental interest \ocite{Haller1224,PabloSolanoExp2019,kao2020creating}, but the highly non-perturbative and strongly correlated nature of the problem makes analytical results scarce. So far, only the sudden interaction quench  starting from the non-interacting ground state, i.e. the Bose-Einstein condensate (BEC), has been theoretically understood \ocite{PhysRevLett.116.070408,SciPostPhys.1.1.001} (although results at special values of the interaction \ocite{PhysRevLett.110.125302} or with a finite number of particles exist \ocite{PhysRevLett.109.115304,PhysRevA.87.053628,SciPostPhys.4.2.011}). In spite of the importance of the result, this protocol has some limitations: first of all, the realization of 1d BECs is difficult due to their instability under thermal fluctuations \ocite{huang2009introduction}. Secondly, a realistic experimental setup is intrinsically inhomogeneous due to the presence of a trapping potential, albeit recently developments allow to engeneer hard-box potentials \ocite{PhysRevX.9.021035,Kwon84}.
Thirdly, the lack of freedom in choosing the initial state results in a narrow variety of steady-states within the attractive regime. 

Generalized Hydrodynamics (GHD) \ocite{PhysRevX.6.041065, Bertini16} is a new powerful toolbox to deal with inhomogeneous integrable models and a new hope in analytically controlling the bound states' production in LL.
Originally introduced to study ballistic transport in integrable systems \ocite{PhysRevX.6.041065, Bertini16,doyonfleagas,PhysRevB.96.115124,PhysRevB.96.081118,collura2018analytic,10.21468/SciPostPhys.8.1.007,Ilievski17,SciPostPhys.3.6.039,10.21468/SciPostPhys.4.6.045, Doyon_2017,doi:10.1063/1.5096892,Bulchandani_2019,PhysRevB.101.035121}, diffusive corrections were subsequently included \ocite{PhysRevLett.121.160603, 10.21468/SciPostPhys.6.4.049,PhysRevLett.122.127202,PhysRevB.98.220303,PhysRevLett.120.164101,10.21468/SciPostPhys.9.5.075}. GHD applications and extensions are now far-reaching, ranging from the study of correlation functions \ocite{10.21468/SciPostPhys.5.5.054,10.21468/SciPostPhysCore.3.2.016}, quantum fluctuations~\ocite{ruggiero2019quantum}, entanglement spreading \ocite{Bertini2018,10.21468/SciPostPhys.7.1.005,alba2017entanglement}, inhomogeneous potentials \ocite{SciPostPhys.2.2.014,PhysRevLett.123.130602,Bastianello_2019,10.21468/SciPostPhys.6.6.070} and integrability-breaking terms \ocite{friedman2019diffusive,PhysRevLett.125.240604,PhysRevB.102.161110,PhysRevLett.126.090602}. Even more, it has been experimentally confirmed \ocite{PhysRevLett.122.090601,malvania2020generalized}.
Of particular interest for the problem at hand is the ability of GHD to describe adiabatically slow modifications of the interaction \ocite{PhysRevLett.123.130602}, provided the underlying integrability structure smoothly changes while varying the coupling. While in LL this is the case within the repulsive and the attractive regime, this is not true any longer when passing from one phase to the other and the state-of-the-art GHD techniques cannot be applied anymore.

In this work, we analytically solve this problem by matching together the hydrodynamic descriptions within the two phases. Our results allow for a complete characterization of the state preparation obtained starting from arbitrary repulsive thermal states (and more general GGEs) and slowly driving the system into the attractive phase.
Our findings can also include the presence of a smooth trapping potential and are feasible for experimental applications, which we briefly discuss.
Our result is checked in the low-density limit against ab-initio microscopic calculations. We also generalize our approach to the case where the interaction is modulated in space, connecting together spatial regions with different interaction signs.
Finally, we discuss how the bound states can be experimentally detected in practice, showing how measurement of the correlated density after a longitudinal trap release can probe the bound states phase-space distribution.

\section{The interacting Bose gas and its GHD} 
The Hamiltonian of the LL model is

\be\label{eq_LLH}
\hat{H}=\int \dd x\,\left\{ \frac{\hbar^2}{2m}\partial_x\hat{\psi}^\dagger\partial_x\hat{\psi}+c\hat{\psi}^\dagger\hat{\psi}^\dagger \hat{\psi}\hat{\psi}+V(x)\hat{\psi}^\dagger\hat{\psi}\right\}\, ,
\ee
\begin{figure}[t!]
\includegraphics[width=0.64\columnwidth]{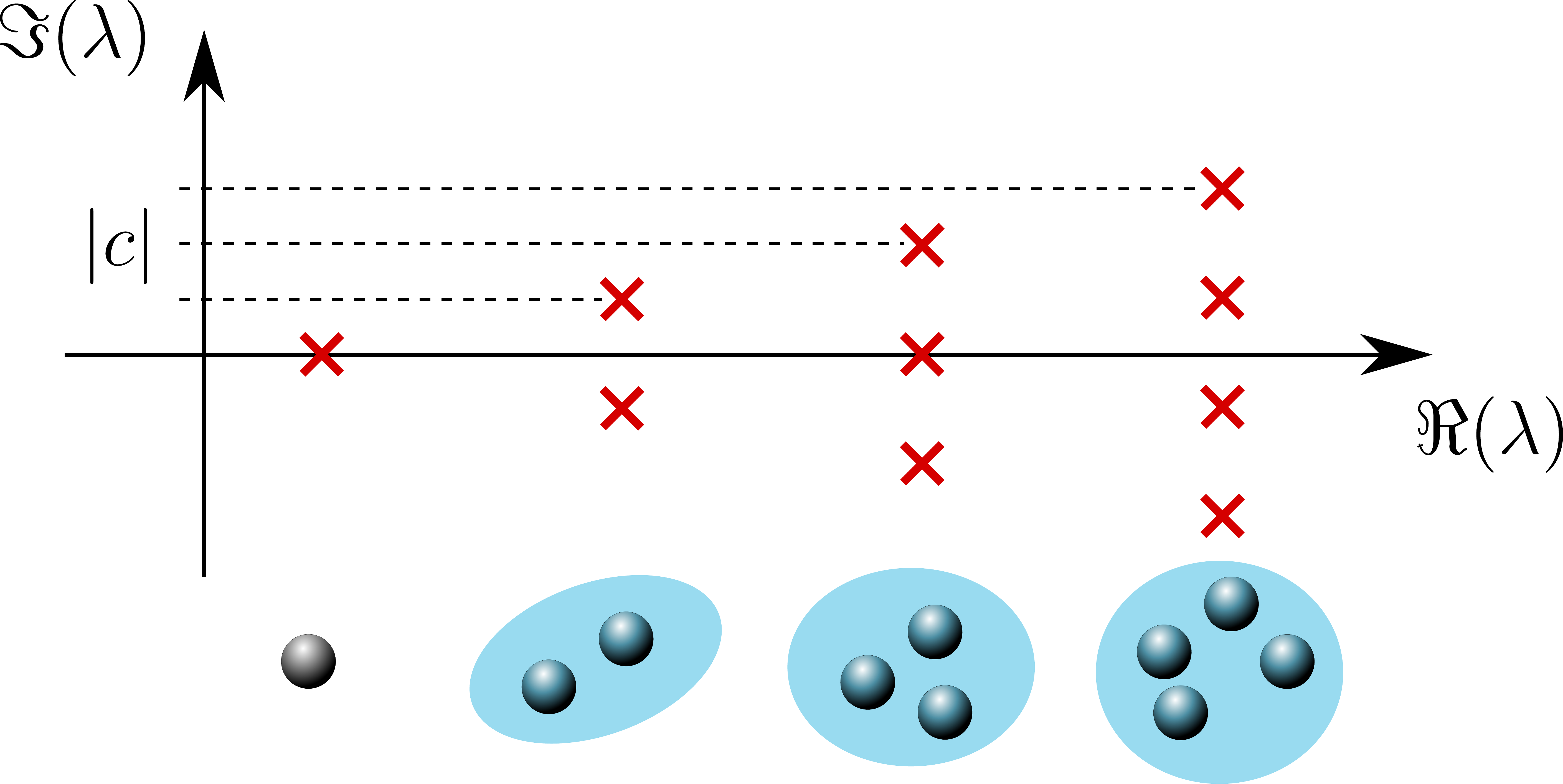}
\includegraphics[width=0.34\columnwidth]{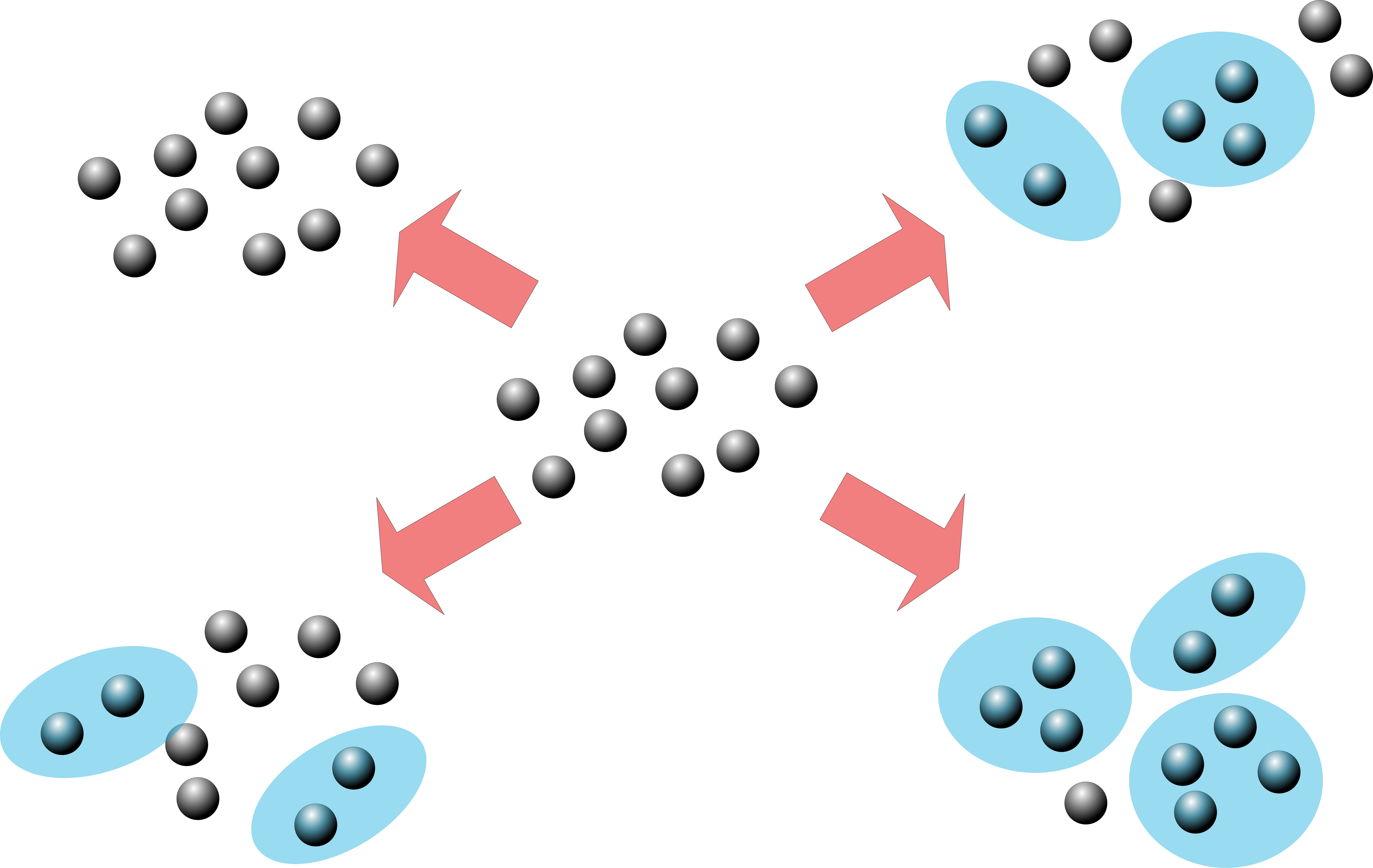}
\caption{\label{fig_1} Left: In the attractive regime, the rapidities of the constituents of a bound state share the same real part, but are shifted in units of $|c|$ along the imaginary axis. Right: As $c\to 0^-$, the bound states are indistinguishable from unbound particles.
}
\end{figure}
where the fields obey standard bosonic commutation relations $[\hat{\psi}(x),\hat{\psi}^\dagger(y)]=\delta(x-y)$, $c$ is the interaction strength and $V(x)$ the external trapping potential. Hereafter, we set our unities such that $\hbar^2/(2m)=1$. 

In the absence of the trap, the model is integrable \ocite{PhysRev.130.1605,PhysRev.130.1616}: the eigenstates of the Hamiltonian and, more generally, of the whole set of local charges can be understood in terms of quasiparticle excitations labeled by a set of quantum numbers $|\{\lambda_i\}_{i=1}^N\rangle$ known as rapidities. For a given (quasi-)local charge \ocite{Ilievski_2016} $\hat{Q}$, the eigenvalue behaves additively $\hat{Q}|\{\lambda_i\}_{i=1}^N\rangle=\sum_{j=1}^N q(\lambda_j)|\{\lambda_i\}_{i=1}^N\rangle$, where the function $q(\lambda)$ is called the charge eigenvalue. At finite volume  these rapidities are quantized according to the Bethe-Takahashi equations \ocite{takahashi2005thermodynamics}, whose solution strongly depends on the sign of the interaction. In the thermodynamic limit and within the repulsive phase $c>0$ the rapidities are real, while for $c<0$ they organize in strings of arbitrary length $j$ \ocite{PhysRevLett.110.257203,franchini2017introduction}: rapidities belonging to a given string share the same real part, but are shifted along the imaginary direction $\lambda_{a}=\lambda-i c(j+1-2a)/2$ with $a\in\{1,j\}$ (see Fig. \ref{fig_1}). Strings can be viewed as bound states of several particles and a string-dependent charge eigenvalue is constructed summing over the constituents of the string $q_j(\lambda)=\sum_{a=1}^j q(\lambda-i c(j+1-2a)/2)$.

The detailed arrangement of rapidities in a given state does not matter in the thermodynamic limit \ocite{PhysRevLett.110.257203,Caux_2016} and the eigenstates are described in terms of root densities $\rho(\lambda)$. In the repulsive case, $L\dd\lambda\rho(\lambda)$ counts how many rapidities in the state are contained in an interval $[\lambda,\lambda+\dd\lambda)$. In the attractive case, infinitely many root densities $\rho_j$ are needed, one for each string species, and describe the occupancy of the real part of the rapidities belonging to the same string. The root densities uniquely identify the thermodynamics of eigenstates and, as such, they are in one-to-one correspondence with the GGEs \ocite{Ilievski_2016_str}.
Let us now allow the system to be weakly inhomogeneous in space and time, but locally integrable. For example, this is the case when an external trap $V(x)$ is introduced and the interaction $c$ becomes space-time dependent. Invoking a separation of scales, one can assume the system locally relaxes to a GGE, which is then slowly evolving: this is the paradigm of GHD \ocite{PhysRevX.6.041065, Bertini16}, which locally describes the system through a space-time dependent root density. The GHD in the presence of a trapping potential and of a space-time dependent interactions is \ocite{PhysRevLett.123.130602} 
\be\label{eq_GHD}
\partial_t \rho_j+\partial_x\left( v_j^\text{eff}\rho_j\right)+\partial_\lambda (F_j^\text{eff}\rho_j)=0\, .
\ee
Qualitatively, this equation describes the evolution of non-interacting particles with phase-space density $\rho_j(t,x,\lambda)$, moving with effective velocity $v^\text{eff}_j$ and experiencing an effective force $F^\text{eff}_j$, where the interactions cause a state dependence of the latter. We wrote the hydrodynamic equations \eqref{eq_GHD} within the attractive phase: the repulsive case is obtained setting $c>0$ in what follows and keeping only the first string $\rho(\lambda)=\rho_1(\lambda)$. The effective velocity and force are defined as
$v^\text{eff}_j=(\partial_\lambda \epsilon_j)^\text{dr}/(\partial_\lambda p_j)^\text{dr}$ and 
\be
F^\text{eff}_j=\frac{\partial_t c f_j^\text{dr}+\partial_x c \Lambda_j^\text{dr}}{(\partial_\lambda p_j)^\text{dr}}-\partial_x V\, ,
\ee
 while 
$\epsilon_j(\lambda)=j \lambda^2-c^2j(j^2-1)/12$ and $p_j(\lambda)=j\lambda$ are respectively the energy and momentum eigenvalues, and 
\begin{eqnarray}\label{eq_f1}
f_j(\lambda)&&=\sum_k\int \frac{\dd\lambda'}{2\pi}\partial_c\Theta_{j,k}(\lambda-\lambda')\rho_k(\lambda')
\\\label{eq_f2}
\Lambda_j(\lambda)&&=\sum_k\int \frac{\dd\lambda'}{2\pi}\partial_c\Theta_{j,k}(\lambda-\lambda')v^\text{eff}_k(\lambda')\rho_k(\lambda')\, .
\end{eqnarray}

The scattering phase $\Theta$ takes into account the interacting nature of the model  $\Theta_{j,k}(\lambda)=(1-\delta_{j,k})\theta_{|j-k|}(\lambda)+2\theta_{|j-k|+2}(\lambda)+...+2\theta_{j+k-2}(\lambda)+\theta_{j+k}(\lambda)$, and $\theta_j(\lambda)=-2\arctan[2\lambda/(jc)]$. Furthermore, the interactions dress the bare quantities according to the linear integral equations \ocite{takahashi2005thermodynamics}

\be\label{eq_dress}
\tau_j^\text{dr}(\lambda)=\tau_j(\lambda)-\sum_k\int \frac{\dd\lambda'}{2\pi}\partial_\lambda\Theta_{j,k}(\lambda-\lambda')\vartheta_k(\lambda')\tau_k^\text{dr}(\lambda')
\ee
where $\vartheta_j=2\pi\rho_j/(\partial_\lambda p_j)^\text{dr}$ is called the filling fraction. 

\section{ Crossing $c=0$} We can finally address our out-of-equilibrium protocol. For the sake of simplicity, let us assume a homogeneous setup and start with a given root density in the repulsive phase, for example a thermal state. The presence of a trap can be included performing the forthcoming analysis at each spatial point. 
Notice that, in the homogeneous case, one can change variable in the GHD equation $t\to c(t)$ and parametrize the root density in terms of the value of the interaction. 

\begin{figure}[t!]
\includegraphics[width=0.99\columnwidth]{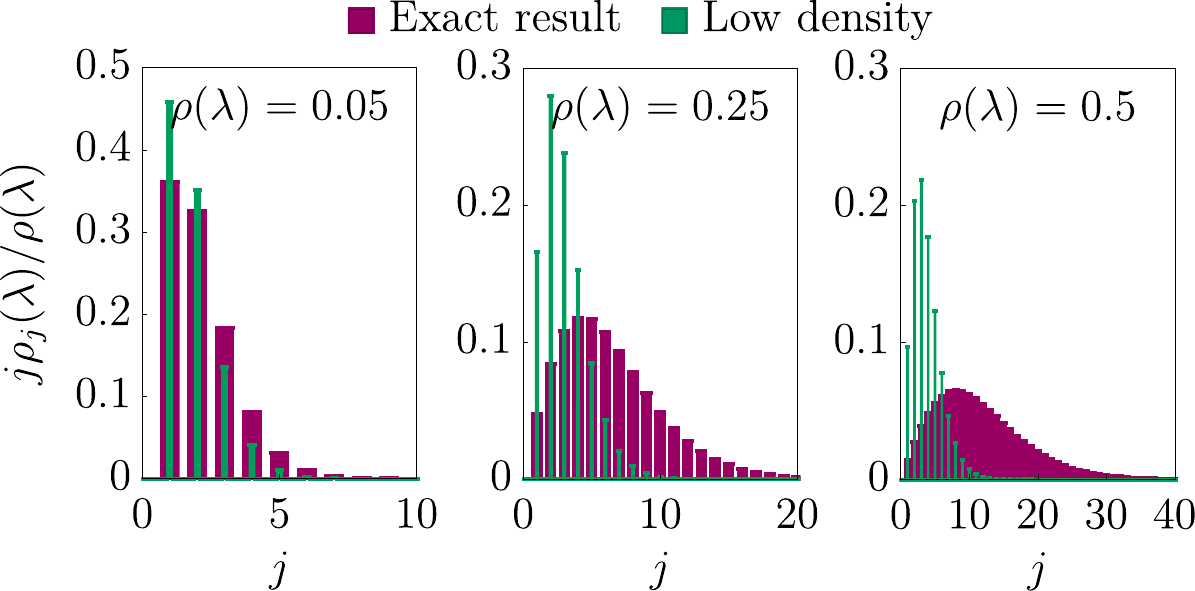}\\ \vspace{0.2cm}
\includegraphics[width=0.99\columnwidth]{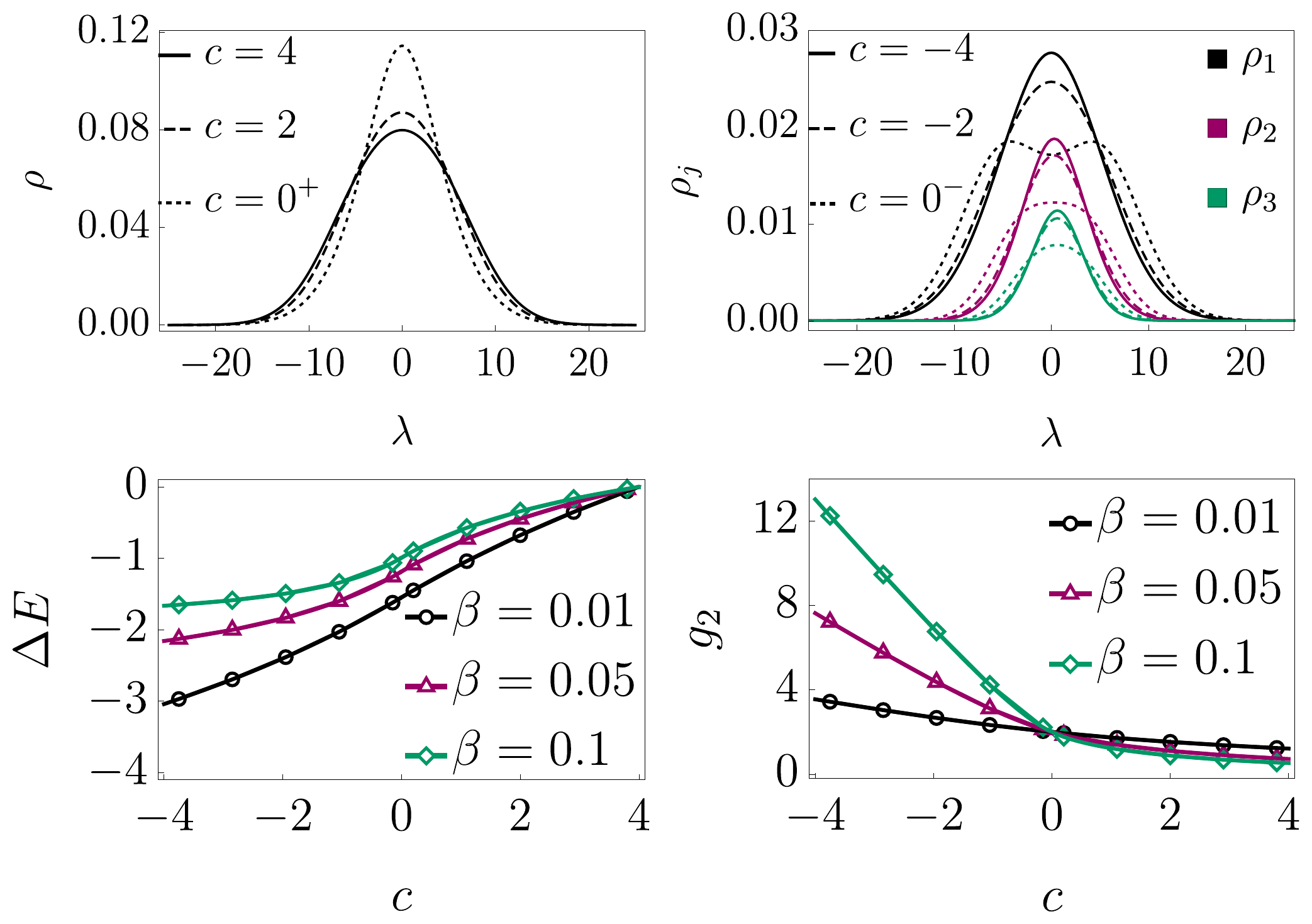}
\caption{\label{fig_2} Top: bound states populations for different choices of $\rho(\lambda)$ at $c=0^-$. The green histogram is obtained using the low-density (incorrect) result $\rho_j=\frac{j}{2\pi}e^{-\omega j}$ (see Eq.\eqref{eq_rho_zd} and the relative discussion).
Middle: evolution of the root densities adiabatically evolving the homogeneous system from $c=4$ to $c=-4$. Notice that, in the homogeneous case, the time evolution can be reparametrized in terms of the interaction $c$. The system is initialized in a thermal state with density $0.5$ and inverse temperature $\beta=0.1$.
Bottom: energy $E$ and $g_2=\langle(\hat{\psi}^\dagger)^2\hat{\psi}^2\rangle/\langle \hat{\psi}^\dagger\hat{\psi}\rangle^2$ evolution for the same protocol, but with different initial temperatures and density $0.5$.
See Appendix \ref{app_fp} for the details of the numerical solution of the GHD equations.
}
\end{figure}

As $c$ is reduced, the particles are compressed together in the rapidity space and the state keeps on evolving until $c=0^+$ is reached (see Fig. \ref{fig_2}).
Here, the endpoint of the adiabatic evolution in the repulsive case must fix the initial conditions for the subsequent evolution in the attractive phase, i.e. determine the set $\{\rho_j\}_{j=1}^\infty$ at $c=0^-$. Of course, the free point $c=0$ can be equally seen as the limit from the weakly repulsive or weakly interacting regime; and the expectation of local observables must be continuous at $c=0$. 
Let us now focus on the conserved charges, whose expectation value is believed to uniquely determine the root densities \ocite{Ilievski_2016_str}.
One has
$\langle\hat{Q}\rangle L^{-1}=\int \dd\lambda\,  q(\lambda)\rho(\lambda)=\sum_j \int \dd\lambda\, q_j(\lambda)\rho_j(\lambda)$, where the system's size $L$ appears because of extensivity. From the charge conservation, we aim to extract the root densities: in order to do so, we first connect the charge eigenvalues in the attractive case with those in the repulsive phase.
The charge eigenvalue is obtained summing over the rapidities belonging to one string, whose imaginary shift vanishes in the $c\to 0^-$ limit
$\lim_{c\to 0^-}q_j(\lambda)=j\lim_{c\to 0^-}q(\lambda)$, 
where the continuity of the charge eigenvalue $q$ is assumed. Finally, it is natural to assume $\lim_{c\to 0^-}q(\lambda)=\lim_{c\to 0^+}q(\lambda)$ since the first string in the attractive case is nothing less than the analytic continuation of the repulsive case. 
The continuity of the charge eigenvalue can be safely assumed for all the local charges, as it is easily checked on the energy and momentum. On the other hand, the attractive phase is expected to feature also quasi-local charges \ocite{Ilievski_2016} due to the presence of bound states. These charges are not expected to be continuous, but their locality properties are associated with the size of the bound states and become less and less local as $c\to 0^-$: hence, these charges are not expected to contribute to the GGE for any state with a finite correlation length $\zeta_\text{cor}$. Indeed, their inclusion in the GGE would induce correlations on a lengthscale $\sim|c|^{-1}\gg \zeta_\text{cor}$.
Thus, invoking the completeness of the local charges, we find the following continuity equation, which we stress is diagonal in the rapidity space
\be\label{eq_continuity}
\rho(\lambda)=\sum_j j\rho_j(\lambda)\, .
\ee
Hence, the charges are unable to fully determine $\{\rho_j\}_{j=1}^\infty$ in the $c\to 0^-$ limit.
The interpretation of this equation is extremely simple: at zero interaction, bound states of $j$ particles with real rapidity $\lambda$ are completely indistinguishable from $j$ unbounded particles with the same rapidity (Fig. \ref{fig_1}). 
For example, this is clear in the two-particle sector, where the wavefunction decays exponentially on a length scale $|c|^{-1}$, i.e. $|\phi(x,y)|\propto e^{-|c||x-y|/2}$ (see Sec. \ref{sec_abinitio}). Precisely, the bound state is indistinguishable from unbound particles when its typical spatial width is much larger than the correlation length of the system. 
The fact that Eq. \eqref{eq_continuity} is diagonal in the rapidity space can be physically motivated as well with the following argument. The interaction acts locally in real space and it is ramped to negative values in an adiabatic fashion, therefore only particles which remain close to each other for an arbitrary long time can bind together. Excitations with different rapidities necessarily have different effective velocities $v^\text{eff}(\lambda)$, therefore are eventually dragged far apart before they can form a bound state.
Eq. \eqref{eq_continuity} can be read in two ways: if the interaction is switched from the attractive to the repulsive regime, $\{\rho_j\}_{j=1}^\infty$ are known and Eq. \eqref{eq_continuity} fully settles $\rho$. In the opposite scenario, $\rho$ is fixed and $\rho_j$ must be determined, a task where Eq. \eqref{eq_continuity} does not suffice.
In order to do this, we revert to the very definition of GGE, i.e. the state that maximizes the entropy under the constraint of fixing the expectation values of all the local integrals of motion. The Yang-Yang (YY) entropy is \ocite{PhysRevLett.110.257203} $S=L\sum_j\int \frac{\dd\lambda}{2\pi}(\partial_\lambda p_j)^\text{dr}[-\vartheta_j\log\vartheta_j-(1-\vartheta_j)\log(1-\vartheta_j)]$ and we will now maximize it under the constrain \eqref{eq_continuity}. 
In the non-interacting limit, the dressing equations are greatly simplified and become diagonal in the rapidity space. Indeed, the derivative of the scattering phase becomes proportional to a Dirac delta
\be\label{eq_theta_csmall}
	\lim_{c\to 0}\partial_\lambda \theta_j(\lambda)=-\text{sign}(c)2\pi\delta(\lambda)\,.
\ee
Since  the same result holds irrespective of $j$, we get
	\be\label{eq_S_delta}
	\lim_{c\to 0}\partial_\lambda \Theta_{jk}(\lambda)=-\text{sign}(c)2\pi \delta(\lambda)[2\min(j,k)-\delta_{jk}].
\ee
As a consequence, in the $c\to 0^-$ limit $(\partial_\lambda p_j)^\text{dr}$ is defined by the following linear equation
\be
(\partial_\lambda p_j)^\text{dr}= j-2\pi\sum_k[2\min(j,k)-\delta_{j,k}]\rho_k(\lambda)\, .
\ee
It is now a simple exercise to maximize the YY entropy with respect to $\rho_j$, using that $\vartheta_j(\lambda)=2\pi\rho_j(\lambda)/(\partial_\lambda p_j)^\text{dr}$. After standard manipulations one gets 
\be\label{eq_effen}
\varepsilon_j(\lambda)=j\omega(\lambda)+\sum_k(2\min(j,k)-\delta_{j,k})\log(1+e^{-\varepsilon_k(\lambda)}),
\ee
where the effective energy parametrizes the filling $\vartheta_j=(1+e^{\varepsilon_j})^{-1}$ and $\omega(\lambda)$ is a $\lambda-$dependent Lagrange multiplier to be determined imposing Eq. \eqref{eq_continuity}.
A similar entropy-maximization strategy has been used in determining the bound-state recombination in the XXZ spin chain affected by a time-dependent magnetic flux \ocite{Bastianello_2019}.
In Fig. \ref{fig_2} we study the bound state formation and provide results for physical observables, initializing the gas in thermal states at $c>0$ and driving it in the attractive regime.

Our approach is valid in the adiabatic regime and it is natural to ask about the allowable time scales. Even though a quantitative analysis of the corrections to the adiabatic result we provided are extremely challenging, we can estimate their validity on the basis of heuristic arguments.

For this discussion, we now go back to dimensionful quantities.
We start by clarifying the nature of the limit $c\to 0^-$: our analysis is built on the assumption that in the non-interacting limit bound states are indistinguishable from unbound particles. More precisely, let us assume that the state has a typical correlation length $\zeta_\text{corr}$. On the other hand, the decay length of the bound states wavefunction is $\zeta_\text{BS}=\frac{\hbar^2}{2m|c|}$.
Whenever $\zeta_\text{BS}\gtrsim \zeta_\text{corr}$,
bound states are effectively indistinguishable from unbound particles. Assuming a linear ramp of the interaction, this sets a timescale $t_\text{BS}$ where mixing between bound states is allowed 
\be
t_\text{BS}\simeq\frac{\hbar^2 }{2m\zeta_\text{corr}}\frac{1}{|\partial_t c|}\, .
\ee
Such time scale must be compared with the assumption of relaxation to a GGE. In integrable models, relaxation is due to dephasing of particles with different velocities \ocite{Sotiriadis_2014}: observing the system at a time $t$, two particles whose initial distance was larger than the correlation length $\zeta_\text{cor}$ will be uncorrelated.
Hence, we estimate the relaxation timescale as
\be
t_\text{rel}\simeq \zeta_\text{corr}/\sqrt{\langle (\Delta v)^2\rangle}\, ,
\ee
with $\langle (\Delta v)^2\rangle$ the variance of the particles' velocity. Requiring $t_\text{rel}\lesssim t_\text{BS}$ we get a crude estimation of the allowed protocol's time scales
\be\label{eq_timescale}
\partial_t c\lesssim \frac{\hbar^2 \sqrt{\langle (\Delta v)^2\rangle}}{2m \zeta_\text{corr}^2}\, .
\ee

	We stress the importance of reaching at $c=0^+$ a finite temperature state for at least two reasons. Firstly, as it is clear from Fig. \ref{fig_2} top panel, the higher the value of the root density $\rho(\lambda)$ is, the larger the produced bound states are. If one would start from the non-interacting homogeneous BEC, the root density diverges at zero rapidity, with the unphysical consequence of creating arbitrary large bound states. The second reason is the validity time scale of the adiabatic approximation: on the non-interacting BEC, the correlation length diverges and the variance of the velocity of the particles goes to zero, hence from Eq. \eqref{eq_timescale} we see that the bound on $\partial_t c$ vanishes. hus, in that particular case, the adiabaticity condition cannot be fulfilled.

Continuous quantum systems are notoriously hard to be simulated \ocite{Schollwoeck2011}, especially out of equilibrium, therefore alternative checks of the validity of our result are extremely important.
In the next section, we provide an ab-initio analysis of the bound states formation in the low density regime, showing how the entropic argument naturally emerges from microscopic calculations and recovering the first term in the low density expansion of \eqref{eq_effen}.
However, before proceeding, we would like to shortly comment on a different sanity check of the finite-density ansatz \eqref{eq_effen}, based on the entropy continuity.
Since crossing the non-interacting point bound states are suddenly available and hence the phase space increased, one can rightfully expect the entropy to increase or, at least, to not decrease.
It turns out that Eq. \eqref{eq_effen} ensures the continuity of the YY entropy passing from $c=0^+$ to $c=0^-$.
Even though the highly non-linear nature of Eq. \eqref{eq_effen} makes the analytical proof of this statement hard, it can be numerically checked with arbitrary precision: this provides a non-trivial check of our results at finite density.
Indeed, since Eq. \eqref{eq_effen} is a maximum, it is also the only choice which guarantees the continuity of the YY entropy: any other solution for $\{\rho_j(\lambda)\}$ that fulfills Eq. \eqref{eq_continuity} would have lowered the entropy of the state, which would have been unphysical.

We also notice that, since the YY entropy is conserved by the GHD equations \ocite{10.21468/SciPostPhys.6.6.070}, it is also conserved during the entire protocol.

\section{Ab-initio analysis of the zero density limit }
\label{sec_abinitio}

The low-density limit is amenable of explicit calculations.
In this section, we derive the ansatz for the formation of the bound states in the zero density limit, using first-principle calculations in the microscopic model.
As we will see, the probability of forming a bound state is completely determined from phase-space arguments and therefore from entropy maximization. Focusing on the zero density limit, we will miss effects of the interaction that are important at finite density, such as dressing. 

The general problem we want to address is the following: let us consider a non-interacting eigenstate labeled by $N$ particles $|\{\lambda_i\}_{i=1}^N\rangle$. Then, we consider another state in the weakly attractive phase featuring some bound states. Let us label it as $|\{\Lambda^j_a\}\rangle$, where $j$ labels the species of the bound state and $a$ is an internal label running on the rapidities $\Lambda$ of the bound states of the same species.
The transition probability $P$ from one state to the other is the overlap squared
\be
P=|\langle \{\lambda_i\}_{i=1}^N|\{\Lambda^j_a\}\rangle|^2\, .
\ee
Our ultimate goal is to compare our calculations with GHD predictions: since the GGE fixes the average occupancy in each rapidity cell $[\lambda,\lambda+\dd\lambda)$, but it is not sensitive to the microscopic arrangements of the rapidities, the probability $P$ must be averaged accordingly.
We already know from the charge eigenvalues that quasiparicles with distinct rapidities cannot bind together. Therefore, let us focus on the case where all the incoming rapidities belong to the same interval $\lambda_i\in [\lambda,\lambda+\dd\lambda)$. The overlap will vanish if the rapidities of the bound states do not belong to the same interval (as we will explicitly see) and, similarly, we are not interested in their exact location on the rapidity axis, but only in their number.
Let $\{n_j\}_{j=1}^\infty$ be the number of bound states of each species at $c=0^-$, then we are interested in the following averaged probability
\begin{multline}\label{av_P}
\bar{P}(\{n_j\})=\sum_{\lambda_i\in[\lambda,\lambda+\dd\lambda)}\sum_{\Lambda_a^j}\frac{|\langle \{\lambda_i\}_{i=1}^N|\{\Lambda^j_a\}\rangle|^2}{[\dd \lambda L/(2\pi)]^N}\Bigg|_{\{n_j\} \text{ fixed}}\, .
\end{multline}
Above, the $\lambda_i$ rapidities are independently summed over the interval $[\lambda,\lambda+\dd\lambda)$.
The prefactor is just the phase space of the rapidities which are quantized as integer multiples of $2\pi/L$, with $L$ the system size.

This object can be explicitly computed in the low density limit in view of very simple considerations. However, as a warm up, it is useful to study first a simple case in detail, namely the probability for $N$ particles to form the largest possible bound state.
\ \\

\subsection{The probability of maximal binding}
For finite $c<0$, the bound state of $N$ particles with rapidity $\Lambda$ and at finite volume in the coordinate representation reads \ocite{franchini2017introduction}

\be
|\Lambda\rangle=\exp\left({i N^{-1}\Lambda\sum_{i=1}^N x_i}\right)\sqrt{\frac{|c|^{N-1}}{L}}\Phi_N(|c|x_1,...,|c|x_N).
\ee
The wavefunction $\Phi$ within the string hypothesis explicitly is
\be\label{eq_SM_bound}
\Phi_N(x_1\le...\le x_N)=\frac{1}{\sqrt{\mathcal{N}_N}}\exp\left[\sum_{a=1}^N x_a (N+1-2a)/2\right]
\ee
and the symmetric extension for other ordering of the coordinates is assumed. However, we will never use the explicit form of $\Phi$, but only that it is translational invariant and fast decaying when the coordinates are stretched far apart.
The incoming state is
\be
|\{\lambda_i\}_{i=1}^N\rangle=\frac{1}{\sqrt{N! L^N}}\sum_P \exp\left({i\sum_a \lambda_{P(a)}x_a}\right)
\ee
where the sum is over the possible permutations $P$ of $N$ elements. Strictly speaking, we assume $\{\lambda_i\}$ to be all different: coinciding rapidities will change the normalization constant, but they are not important after the averaging. One needs to compute the overlap
\begin{multline}\label{eq_20}
\langle\Lambda|\{\lambda_i\} \rangle=\frac{\sqrt{|c|^{N-1}}}{\sqrt{N! L^{N+1}}}\sum_P \int \dd^N x\, \Phi(|c|x_1,...,|c|x_N) \times\\
\exp\left({i\sum_a \lambda_{P(a)} x_a-\frac{i\Lambda}{N} \sum_a x_a}\right)\, .
\end{multline}
From this overlap, straightforward albeit tedious computations allow one to access the averaged probability Eq. \eqref{av_P}. For arbitrary values of the interactions, $\bar{P}$ is a complicate object, but it is greatly simplified in the $c\to 0^-$ limit: in this case, the overlap $\langle\Lambda|\{\lambda_i\} \rangle$ gets extremely peaked in the rapidity space and non vanishing only for $|\Lambda-\lambda_i|\lesssim |c|$.
We leave the detailed computations to Appendix \ref{app_ov_calc} and simply quote the final expression
\be\label{eq_24}
\bar{P}=\left[\frac{1}{N!}\left(\frac{\dd\lambda L}{2\pi}\right)^N\right]^{-1} \frac{\dd\lambda L N}{2\pi}\, .
\ee
This result can be now interpreted in terms of phase-space densities. The quantity $\dd\lambda L N/(2\pi)$ is the number of possible rearrangements of the momentum of the bound state of $N$ particles in the interval $[\lambda,\lambda+\dd\lambda)$. Indeed, since $\Lambda=N^{-1}\sum_a \lambda_a$ and $\lambda_a$ are quantized in units of $2\pi L^{-1}$, $\Lambda$ is therefore quantized in units of $2\pi L^{-1}N^{-1}$. On the other hand, $\frac{1}{N!}\left(\frac{\dd\lambda L}{2\pi}\right)^N$ is the phase space of the incoming states (the $N!$ term keeps into account the fact that particles are indistinguishable).
The phase-space density result suggests that the probability in the general case can be determined on the basis of simple arguments. This is indeed the case.

\subsection{The probability of the generic transition}
We can now go back to the problem of computing the transition probability to an arbitrary state, in the $c\to 0$ limit. In order to do this, we use some assumptions which, based on our previous calculations, are expected to be valid in the zero-density limit.
Let us consider the generic overlap $\langle\{\lambda_i\}|\{\Lambda_a^j\}\rangle$: this overlap can be divided into a product of Kronecker deltas enforcing the conservation laws of the momenta times a smooth part $F$
\be
|\langle\{\lambda_i\}|\{\Lambda_a^j\}\rangle|^2=[\text{conservation law}]\times F(\lambda_1,...,\lambda_N)\, .
\ee
 In the maximal binding calculation the conservation law was extremely simple $\delta_{\Lambda,N^{-1}\sum_a\lambda_a}$. Instead, in the general case, the constraint is a complicated product of Kronecker deltas (and sum of this product over rapidity permutations). The set of rapidities $\{\lambda_i\}$ is partitioned into groups of rapidities, where the center of mass of each group is enforced to be equal to the rapidity of a certain bound state.
However, there is no need to make this constraint explicit. 
As we commented in the explicit computation in the maximal binding case, in the $c\to 0^-$ limit the smooth part of the overlap becomes extremely peaked around the rapidity of the bound state: we assume it is the case in the generic overlap as well, therefore $F$ is peaked for the rapidities $\lambda_i$ close to the rapidity of the associated bound state.
This observation allows one to write
\begin{multline}
\sum_{\lambda_i\in [\lambda,\lambda+\dd\lambda)}|\langle\{\lambda_i\}|\{\Lambda_a^j\}\rangle|^2\to \\
\sum_{\lambda_i}[\text{conservation law}]\times F(\lambda_1,...,\lambda_N)=N!\, .
\end{multline}
The second sum is unconstrained. Since the rapidities of the bound state belong to the interval $[\lambda,\lambda+\dd\lambda)$, the function $F$ is zero whenever one of the rapidities $\lambda_i$ lays outside of the interval. The unconstrained sum is then equal to $N!$, because of the completeness of the $|\{\lambda_i\}\rangle$ states.
In order to get the averaged probability $\bar{P}$, we need now to sum over the possible positions of the bound states within the interval $[\lambda,\lambda+\dd\lambda)$. The rapidity of a bound state of species $j$ is quantized in units of $2\pi/(Lj)$, therefore we get
\begin{multline}
\sum_{\Lambda_a^j\in [\lambda,\lambda+\dd\lambda)}\sum_{\lambda_i\in [\lambda,\lambda+\dd\lambda)}|\langle\{\lambda_i\}|\{\Lambda_a^j\}\rangle|^2=\\=\sum_{\Lambda_a^j\in [\lambda,\lambda+\dd\lambda)} N!=
 N!\prod_j \left[\frac{1}{n_j!}\left(\frac{\dd\lambda L j}{2\pi}\right)^{n_j}\right]\, .
\end{multline}
Above, the $n_j!$ terms account for the indistinguishability of the bound states and, in the zero density limit, we quantize the rapidities of the bound states independently.
This simple analysis gives us the following simple averaged probability
\be\label{eq_P}
\bar{P}(\{n_j\})=\left[\frac{1}{N!}\left(\frac{\dd\lambda L}{2\pi }\right)^N\right]^{-1}\prod_j\left[\frac{1}{n_j!}\left(\frac{\dd\lambda L j}{2\pi}\right)^{n_j}\right]+...\, ,
\ee
which is of course consistent with the maximal binding probability previously derived. We notice that the averaged probability is completely determined in terms of the phase space. Above, further corrections are present: indeed, the probability $\bar{P}$ is not correctly normalized. One can simply observe, for example, that the configuration $n_1=N$ and $n_{j>1}=0$ already saturates the probability $\bar{P}(n_1=N,n_2=0,...)=1$.
The reason is the following: the argument we provided only captures the leading behavior in the thermodynamic limit of a given configuration $\{n_j\}$. Indeed, if one considers the power counting in $L$ factors finds $L^{\sum_j n_j-N}$. Using the constraint $N=\sum_j jn_j$, the power counting can be rewritten as $L^{-\sum_j (j-1) n_j}$: therefore, strictly speaking, in the $L\to \infty$ limit only the configuration $n_1=N$ survives, while other configurations vanish and the probability is correctly normalized. Hence, at finite $L$, the probability of each configuration has non-trivial corrections to subleading orders in $L$.

\subsection{The averaged population in the low-density limit}
We finally use the probability $\bar{P}$ \eqref{eq_P} to compute the average bound-state population.
Since $\bar{P}$ captures only the leading order in $L^{-1}$, the resulting expectation values $\langle n_j\rangle$ will be valid at the leading order in the zero density limit as well.

Let us consider $\bar{P}$ \eqref{eq_P} in the limit of large occupation numbers $n_j$. In this case, one can use the Stirling approximation and write
\be
\bar{P}\propto \exp\left[\sum_j n_j\log\left(\frac{\dd\lambda L j}{2\pi}\right)-n_j\log n_j+n_j\right]\, .
\ee
This expression can be immediately compared with the low-density regime of the YY entropy. Indeed, if one identifies  $L\dd\lambda \rho_j= n_j$ and $L\dd\lambda \rho=N$, $\bar{P}$ can be expressed as
\be
\bar{P}\propto \exp\left[L\dd\lambda\sum_j  \rho_j\log\left(\partial_\lambda p_j\right)-\rho_j \log (\rho_j)+\rho_j\right]\, ,
\ee
where we used that $\partial_\lambda p_j=j/(2\pi)$.
The argument in the exponential is nothing else than the leading order in the $\rho_j\to 0$ limit of the YY entropy. In the $L\to\infty$ limit, the probability is peaked around its maximum and therefore the expectation values are determined by the saddle point, namely entropy maximization. While doing so, one should take into account the constraint $\rho=\sum_j j\rho_j$ that is trivially derived from $N=\sum_j j n_j$.
The entropy maximization gives the following simple result
\be\label{eq_rho_zd}
\rho_j=\frac{j}{2\pi}e^{-\omega j}\, ,
\ee
with $\omega$ a Lagrange multiplier. Then, $\omega$ is fixed by $\rho=\sum_j j\rho_j$. As expected, this is the low-density limit of the GHD result.

\section{Generalization to spatial inhomogeneities} Our result can be promptly extended to spatially inhomogeneous interactions. 
Besides the theoretical interest, this generalization is motivated by the recent advances in the experimental engineering of inhomogeneous interactions \ocite{PhysRevLett.115.155301}.
Furthermore, in the setup of Ref. \ocite{Haller1224} the gradient in the magnetic field used to counterbalance the gravitational force causes a weakly inhomogeneous interaction \ocite{NagerlLandini2020pvt}.
Let us consider $c(x\lessgtr 0)\lessgtr 0$ to be a smoothly inhomogeneous function constant in time. The strings flowing from the attractive to the repulsive region $(\lambda>0)$ unbind, while particles can form bound states when traveling in the other direction $(\lambda<0)$. In this case, rather than the continuity of the charges one must impose the continuity of the current associated with the latter. This requirement leaves some freedom in choosing the bound-state populations, which can be again determined by maximum entropy considerations. However, in this case, the entropy rate must be maximized: computing $\partial_t S$ with the help of the GHD equations, one finds that this rate is completely determined by the root densities at $x=0$.
A detailed derivation is provided thereafter, leading to the same equations as before, namely Eqs. \eqref{eq_continuity} and \eqref{eq_effen}.

Within the $x>0$ and $x<0$ regions, the Eulerian dynamics is entirely governed by the GHD equations and, similarly to the time-dependent protocol, one has to find the proper boundary conditions at the transition point.
First, we find the analogue of charge-conservation, which can be easily understood to be the continuity of currents. Indeed, any discontinuity of a current would imply a divergent growth of the associated charge density, that is of course unphysical.
The exact expression for currents has already been proposed in the original papers on GHD \ocite{PhysRevX.6.041065, Bertini16} and it reads
\be
\langle\hat{J}\rangle=\sum_j \int \dd \lambda \, v^\text{eff}_j (\lambda) q_j(\lambda)\rho_j(\lambda)\, ,
\ee
where $q_j$ is the charge eigenvalue of the charge $\hat{Q}$ associated with the current.

Assuming the
analiticity of the charge eigenvalues and their completeness (together with $\lim_{c\to 0^-}q_j(\lambda)=j\lim_{c\to 0^+}q(\lambda)$), one gets a continuity equation. Thanks to the fact that, at $c\to 0$, the dressing acts diagonally in the rapidity space, from the very definition of the effective velocity one can easily show that, both in the repulsive and attractive regime, it holds $v^\text{eff}(\lambda)=v^\text{eff}_j(\lambda)=2\lambda$.
This further simplifies the continuity equation obtained from the currents which, in the end, is identical to the time-dependent case
\be\label{eq_j_continuity}
\rho(\lambda)=\sum_j j  \rho_j(\lambda)\, .
\ee

Similarly to the charge conservation in the time-dependent case, Eq. \eqref{eq_j_continuity} does not completely fix the boundary conditions, since it allows a possible rearrangement of the bound states.
More specifically,
the current flowing into the junction is of course fixed by the left and right bulks, while the out-going current must be found. The notion of in-going and out-going is determined by the sign of $v^\text{eff}(\lambda)=v_j^\text{eff}(\lambda)=2\lambda$.
In order to unambiguously determine the bound state recombination, we consider the entropy once again.

Within the inhomogeneous setup, rather than considering the YY entropy, one should focus on its growth.
Let us consider $\partial_t S=\partial_t S_{x<0}+\partial_t S_{x>0}$, where $S_{x\lessgtr 0}$ is the Yang-Yang entropy in the left and right halves of the system, respectively.
The GHD equations have been proved to conserve the entropy \ocite{10.21468/SciPostPhys.6.6.070}, however this is true only in the absence of boundary terms (see also Ref. \ocite{Bastianello_2019}).
Indeed, let us consider the Yang-Yang entropy within the attractive regime (i.e. in the region $x<0$) and compute its time derivative, using the GHD equations one straightforwardly obtains
\begin{multline}
\partial_t S_{x<0}=\partial_t\left(\int_{-\infty}^0 \dd x\int \frac{\dd\lambda}{2\pi} \sum_j (\partial_\lambda p_j)^\text{dr}\eta(\vartheta_j)\right)=\\
-\sum_j \int_{-\infty}^0 \dd x\int \frac{\dd\lambda}{2\pi}\,\Big[ \partial_x(v^\text{eff}_j(\partial_\lambda p_j)^\text{dr} \eta(\vartheta_j))+\\
\partial_{\lambda}\left(F^\text{eff}_j(\partial_\lambda p_j)^\text{dr}\eta(\vartheta_j)\right)\Big]\, ,
\end{multline}
where $\eta(x)=-x\log x-(1-x)\log(1-x)$. Since we integrate exact differentials, only boundary terms matter. In the hypothesis that the filling vanishes at large rapidities and for $x\to-\infty$, we get a non trivial contribution only from the boundary at $x=0$
\be\label{S_eq_entrogrowth}
\partial_t S_{x<0}=-\sum_j \int \dd\lambda\, v^\text{eff}_j\rho_j^t \eta(\vartheta_j)\Bigg|_{x=0}\, .
\ee
A similar conclusion holds for $S_{x>0}$.
Now, we are left with the problem of maximizing the entropy rate with the constrain \eqref{eq_j_continuity}. Of course, we are considering the $c\to 0$ limit, hence the dressing is diagonal in the rapidity space and $v_j^\text{eff}=2\lambda$: using this identity in \eqref{S_eq_entrogrowth}, we obtain that the integrand in the entropy growth is, apart from the factor $2\lambda$, exatcly the YY entropy we maximized in the time-dependent case. Since the dressing acts diagonally in the rapidity space, the $2\lambda$ prefactor is ineffective in the entropy maximization. Besides, the continuity equation \eqref{eq_j_continuity} is formally the same as what we had in the time-dependent case. Hence, the entropy maximization leads to exactly the same non-linear equations (4-5).

\section{Bound states' detection in experiments}

We expect our results to be applicable to the state-of-the-art experimental techniques.
In Refs. \ocite{Haller1224} and \ocite{kao2020creating} Cesium and Dysprosium atoms  respectively were trapped in 1d optical traps and the interaction manipulated acting on a Feshbach resonance \ocite{PhysRevLett.91.163201}. In particular, by gently tuning the magnetic field, the whole range from weakly repulsive to strongly attractive interactions can be continuously explored. Ref. \ocite{Haller1224} focused on sudden interaction changes, while Ref. \ocite{kao2020creating} implemented an adiabatic protocol.
The initial state $c>0$ is expected to be thermal and its temperature can be estimated by measuring the mean kinetic energy trough momentum-space imaging. With the same method, the kinetic energy can be probed at the end of the protocol and compared with the GHD result.
Advances in atom-chip setups \ocite{bederson2003advances} could lead to even more interesting measurements, given the possibility of real-space density's profile imaging. Combining the latter with a longitudinal trap release, the rapidity-dependent root densities of the bound states can be reconstructed from the full-counting statistics of the density fluctuations, as we now discuss.

Let us imagine a 1d interacting Bose gas confined in an elongated trap, with homogeneous and time-independent interaction. In Ref. \ocite{10.21468/SciPostPhys.6.6.070} it has been pointed out that, within the repulsive phase, the gas expansion following a longitudinal trap release (but maintaining the transverse confinement), allows one to reconstruct the rapidity-dependent root density, integrated in space. The same method can be used to detect the population of the bound states within the attractive phase through correlated density measurements of the expanding cloud.
First, we quickly recap the measurement proposed in Ref. \ocite{10.21468/SciPostPhys.6.6.070}, then we move to discuss the attractive regime. This method has been used, for example, in Ref. \ocite{malvania2020generalized} for extracting the root density from experimental data.
We consider $c>0$ and imagine the longitudinal trap is released, but the transverse trap is kept in place retaining the 1d geometry of the gas. The free expansion is determined by the GHD equations $\partial_t\rho+\partial_x(v^\text{eff}\rho)=0$: while the cloud expands, its local density decays in time due to the ballistic spreading $\sim t^{-1}$ hence, after a certain large time $t_0$ after the trap release, dressing effects in the velocity can be neglected $v^\text{eff}(\lambda)\simeq v(\lambda)=2\lambda$.
In this regime, the solution to the GHD equations amounts to a free expansion
\be
\rho(t,\lambda,x)=\int \dd y\, \delta\left(y-x+2\lambda (t-t_0)\right)\rho(t_0,\lambda,y)\, ,
\ee
Now, let us consider the density profile and compute $\langle \hat{\psi}^\dagger(x)\hat{\psi}(x)\rangle=\int \dd\lambda\, \rho(t,\lambda,x)$. Using the expression above, one finds
\be
\langle \hat{\psi}^\dagger(x)\hat{\psi}(x)\rangle=\frac{1}{2 (t-t_0)}\int \dd y\,  \rho\left(t_0,\frac{x-y}{2(t-t_0)},y\right)\, .
\ee

We now perform a further approximation and take $t\gg t_0$ and observe the cloud at positions $x$ much larger than the cloud's size at time $t_0$ (in terms of adimensional quantities $\frac{x-y}{2(t-t_0)}\times\left[\partial_\lambda \rho/\rho\right]_{\lambda=\frac{x-y}{2(t-t_0)}}\ll 1$). 
Within this approximation, one finds $\langle \hat{\psi}^\dagger(x)\hat{\psi}(x)\rangle=\frac{1}{2 t}\int \dd y\,  \rho\left(t_0,\frac{x}{2t},y\right)$, lastly we notice that the ghd equations $\partial_t\rho+\partial_x (v^\text{eff} \rho)=0$ implies that $\int \dd y\,  \rho\left(t,\lambda,y\right)$ is conserved during the time evolution, hence we can replace $t_0\to t$ in the expression for the density profile and finally get
\be
\langle \hat{\psi}^\dagger(x)\hat{\psi}(x)\rangle=\frac{1}{2 t}\int \dd y\,  \rho\left(0,\frac{x}{2t},y\right)\, .
\ee
Hence, measuring the density profile of the expanding cloud, one can measure $\int\dd y\, \rho(0,\lambda,y)$ as a function of the rapidity $\lambda$ and realize a spectroscopy of the root density.

A similar reasoning can be applied if the gas is in the attractive phase, although measurements of the density profile do not allow to discern among the bound states. 
Indeed, applying the same reasoning we outlined in the repulsive case, at $c<0$ one finds

\be
\langle \hat{\psi}^\dagger(x)\hat{\psi}(x)\rangle=\frac{1}{2 t}\sum_j\int \dd y\,  j\rho_j\left(0,\frac{x}{2t},y\right)\, .
\ee
This single observable cannot distinguish the bound states and further measurements are needed. Physically, bound states are cluster of correlated particles which travel together. Hence, the correlated densities $\mathcal{O}_\ell(x)=[\hat{\psi}^\dagger(x)]^\ell[\hat{\psi}(x)]^\ell$ are the natural candidates to probe the bound states' population.
Within the GHD perspective, these observables can be expressed in terms of a functional of the root densities at the same position $\langle\mathcal{O}_\ell(x)\rangle=O_\ell[\rho_j(x)]$. At finite density, such a functional is complicated and in general not known. However, we can use the fact that in the large time limit after the trap release, the local density of particles is small and attempt a linear expansion
\be\label{eq_SM_Ol_texp}
\langle \mathcal{O}_\ell(x)\rangle\simeq \sum_j \int \dd\lambda\, \mathcal{C}_j^\ell(\lambda) \rho_j(t,x,\lambda)\, .
\ee
Since the coefficients $\mathcal{C}_j^\ell$ describe the low density limit of the expectation values $\langle \mathcal{O}_\ell\rangle$, they can be explicitly computed in the finite-particle sector.
More precisely, let $|j,\lambda\rangle$ be the bound state of $j$ constituents and rapidity $\lambda$, with $L$ the system's length
\begin{multline}
|j,\lambda\rangle=\sqrt{\frac{|c|^{j-1}}{j!L\mathcal{N}_j}}\int \dd^j x \, \exp\left[i\frac{\lambda}{j}\sum_{a=1}^j x_a \right]\times\\
\Phi_j(|c|x_1,...,|c|x_j)\hat{\psi}^\dagger(x_1)...\hat{\psi}^\dagger(x_j)|0\rangle\, .
\end{multline}
The wavefunction $\Phi$ is reported in Eq. \eqref{eq_SM_bound}. The normalization $\mathcal{N}_j$ in Eq. \eqref{eq_SM_bound} can be fixed imposing $\int_{x_1=0}\dd^{j-1}x\, |\Phi_j(x_1,...,x_j)|^2=1$.

Let us now focus on computing the expectation value of $\mathcal{O}_\ell(x)$ on this state. Thanks to translational invariance, we can consider $\mathcal{O}_\ell(x=0)$ and one readily finds
\be
\langle j,\lambda|\mathcal{O}_\ell(0)|j,\lambda\rangle=
\frac{\ell!}{  L}\binom{j}{\ell}\int\dd^{j-\ell}x |\Phi_j(x_1,...,x_{j-\ell},0,..0)|^2
\ee
for $j\ge \ell$ and zero otherwise.
Comparing the above with Eq. \eqref{eq_SM_Ol_texp} and using that the state $|j,\lambda\rangle$ is represented by a root density $\rho_{j'}(\lambda')=L^{-1}\delta_{j,j'}\delta(\lambda-\lambda')$, one can determine the coefficients $\mathcal{C}_j^\ell(\lambda)$:
\be
\mathcal{C}_j^\ell(\lambda)=\ell!\binom{\ell}{j}\int\dd^{j-\ell}x |\Phi_j(x_1,...,x_{j-\ell},0,..0)|^2
\ee
for $j\ge \ell$ and zero otherwise.
Notice that, due to Galilean invariance, $\mathcal{C}_j^\ell(\lambda)$ is actually $\lambda-$independent.
Once these coefficients have been computed, one can analyze the expanding cloud, lastly determining $\int \dd x\, \rho_j(t,\lambda,x)$ as a function of $\lambda$.

The computation of $\mathcal{C}_j^\ell$ requires performing a finite-dimensional integral in at most $j$ coordinates of a simple wavefunction \eqref{eq_SM_bound}. Albeit tedious, this task can be straightforwardly performed, especially in the case where only the first strings are populated and one can truncate the sum \eqref{eq_SM_Ol_texp}.

The local correlated density $[\hat{\psi}^\dagger(x)]^\ell[\hat{\psi}(x)]^\ell$ can be difficult to be directly measured in experiment. However, the same strategy can be used on the moments of the particle numbers in an interval. Let us define $\hat{N}_\Delta(x)=\int_{x-\Delta/2}^{x+\Delta/2} \dd x'\, \hat{\psi}^\dagger(x')\hat{\psi}(x')$, i.e. the particle number operator in an interval of length $\Delta$ centered around $x$.
In the long time limit after the trap release, one can write
\be
\langle [\hat{N}_\Delta(x)]^\ell\rangle=\frac{1}{2t}\sum_j \int \dd y\, \tilde{\mathcal{C}}_j^\ell \rho_j\left(0,\frac{x}{2t},y\right)\, ,
\ee
where the computation of the coefficients $\tilde{\mathcal{C}}_j^\ell$ is a trivial generalization of the strategy that led to determine $\mathcal{C}_j^\ell$.

\section{Conclusions and outlook} We analytically predict the bound states' formation in the 1d interacting Bose gas undergoing adiabatic interaction changes from the repulsive to the attractive regime.  
Our exact results are valid in the thermodynamic limit and also when correlations are strong and inaccessible to perturbation theory. We considered generic initial thermal states and more generally GGEs: this flexibility allows to greatly control the attractive phase with immediate applications to state-preparation. Our findings are experimentally accessible,
but also provide prospects for further developments in inhomogeneous 1d systems.
For example, inhomogeneous spin chains can arguably be studied with similar methods and the consequences of bound states' recombination on transport problem addressed \ocite{Biella2019}. 
The experimental setup of Ref. \ocite{Gring1318} represents a major theoretical challenge, with a cold-atom realization of the famous sine Gordon (SG) model, describing the phase interference between two coupled 1d atom tubes. The intrinsic inhomogeneity induced by the experimental setup causes a smooth space dependence on the SG interaction, which strongly affects the local spectrum of the theory, and causes binding and unbinding the topological excitations of the phase.
Our findings are a first step towards the solution of this very interesting, but difficult, problem.
Future applications to classical systems are thriving to be addressed as well: in the semiclassical limit, the 1d Bose gas reduces to the 1d Non-Linear Schr\"odinger equation \ocite{De_Luca_2016}.
This classical correspondence allowed for several numerical benchmarks of predictions dragged from integrability \ocite{10.21468/SciPostPhys.9.1.002,PhysRevB.102.161110,PhysRevLett.125.240604,PhysRevA.99.023605} and could offer a numerical confirmation of our method.

\section*{Acknowledgements}
The authors acknowledge support from the European Research Council (ERC) under ERC Advanced grant 743032 DYNAMINT. We are grateful to A. De Luca, J. De Nardis and F. H. L. Essler for useful discussions.
We are indebted to H.-C. N\"agerl, M. Landini for a detailed discussion of the experimental apparatus of Ref. \ocite{Haller1224} and to B. Lev for pointing out Ref. \ocite{kao2020creating}.

\appendix

\section{Numerical solution of the GHD equation and observables of interest}
\label{app_fp}

In this appendix, we provide additional details concerning the numerical solution of the GHD equations and how the plots in Fig. \ref{fig_2} have been obtained.
For the sake of simplicity, we considered a spatially homogeneous system, but the method is readily generalized to inhomogeneous systems.

We initialize the state in a thermal ensemble at a finite temperature and $c>0$, whose filling function is determined by the following integral equation \ocite{takahashi2005thermodynamics}
\be
\varepsilon(\lambda)=\beta[\epsilon(\lambda)-\mu]+\int \frac{\text{d}\lambda'}{2\pi}\partial_{\lambda}\theta_1(\lambda-\lambda')\log\left(1+e^{-\varepsilon(\lambda')}\right) \, .
\ee
Above, $\mu$ is a chemical potential fixed by the density of particles.
This equation, as well as the dressing \eqref{eq_dress}, is discretized on a finite uniform grid $\{\lambda_i\}_i$ with a lattice step $\Delta \lambda$.
\be
\varepsilon(\lambda_i)=\beta[\epsilon(\lambda_i)-\mu]+\sum_{i'} \frac{1}{2\pi}[\partial_{\lambda}\theta_1]_{i,i'}\log\left(1+e^{-\varepsilon(\lambda_{i'})}\right) \, .
\ee
with $[\partial_{\lambda}\theta_1]_{i,i'}$ the discretized kernel.
One could naively set $[\partial_{\lambda}\theta_1]_{i,i'}=\Delta\lambda \partial_{\lambda}\theta_1(\lambda_i-\lambda_{i'})$, but such a  discretization does not work at small $c$ in view of Eq. \eqref{eq_theta_csmall}, hence we rather define
\begin{multline}
[\partial_{\lambda}\theta_1]_{i,i'}=\int_{\lambda_{i'}-\Delta/2}^{\lambda_{i'}+\Delta/2} \dd\lambda'\, \partial_{\lambda}\theta_1(\lambda-\lambda')=\\
2\arctan\left[\frac{2(\lambda'-\lambda_i)}{c}\right]\Bigg|^{\lambda'=\lambda_{i'}+\Delta/2}_{\lambda'=\lambda_{i'}-\Delta/2}
\end{multline}
A similar discretization strategy must be employed for the dressing within the attractive phase and for the force terms (\ref{eq_f1}-\ref{eq_f2}).

The initial state is then evolved with the GHD equations according to the method of characteristics used in Ref. \ocite{PhysRevLett.123.130602}, which implements the GHD equation as infinitesimal and inhomogeneous translations of the filling function in the phase space. Lastly, once $c=0^+$ is reached, the evolution is continued within the attractive phase solving Eq. \eqref{eq_effen}.

During the evolution, we mainly focus on two physically motivated observables, namely the total energy $E$ and the correlated density $g_2=\langle (\hat{\psi}^\dagger)^2\hat{\psi}^2\rangle/(\langle\hat{\psi}^\dagger\hat{\psi}\rangle)^2$. In terms of the root densities and filling fractions, these observables are 

\be
E=\sum_{j}\int \dd \lambda\, \epsilon_j(\lambda)\rho_j(\lambda)\,, \hspace{1pc}
\langle \hat{\psi}^\dagger\hat{\psi}\rangle=\sum_{j}\int \dd \lambda\, j\rho_j(\lambda)
\ee
and \ocite{PhysRevLett.116.070408,SciPostPhys.1.1.001}
\begin{multline}
\langle(\hat{\psi}^\dagger)^2\hat{\psi}^2\rangle=-\sum_j\int \dd\lambda\frac{c}{6}j(j^2-1)\rho_j(\lambda)+\\\sum_j\int\frac{\dd\lambda}{\pi}j\lambda\vartheta_j(\lambda) f_j^\text{dr}(\lambda)\, .
\end{multline}
Both observables are reported in the attractive phase, while the repulsive one is obtained retaining only the first string $j=1$ and of course $c>0$.

\section{From the overlap to Eq. \eqref{eq_24}}
\label{app_ov_calc}

In this appendix, we present the detailed calculations that from the overlap \eqref{eq_20} bring one to Eq. \eqref{eq_24}.

The symmetry of the wavefunction $\Phi$ under global translations allows one to integrate the center of mass $N^{-1}\sum_a x_a$  in Eq. \eqref{eq_20} and the oscillating phases impose the momentum constraint $\Lambda=N^{-1}\sum_a \lambda_a$.
Importantly, we work at large but finite volume, hence conservation laws are not enforced through Dirac deltas, but Kronecker deltas and, of course, $L$ factors. Rather than aiming for a brute-force computation of the integral, it is more convenient to keep the formal integral representation and plug it directly into Eq. \eqref{av_P}.
The summation over the $\Lambda$ states is trivially performed, since the only non zero contribution is when $\Lambda=N^{-1}\sum_a\lambda_a$. Then, we use the fact that since we are summing over all the rapidities and coordinates, the result is invariant under permutations. Therefore, we can pick a single arrangement of coordinates and rapidities and introduce a prefactor $(N!)^2$ to keep into account the double summation over the rapidities, which gives
\begin{widetext}
\be
\bar{P}=\left(\frac{\dd\lambda L}{2\pi}\right)^{-N} \frac{N!|c|^{N-1} }{L^{N+1}}\sum_{\lambda_i\in[\lambda,\lambda+\dd\lambda)}
\int \dd^N x \int \dd^N x'\, \Phi(|c|x_1,...,|c|x_N)\Phi(|c|x'_1,...,|c|x'_N) e^{i\sum_a (\lambda_{a}-N^{-1}\sum_{a'}\lambda_{a'}) (x_a-x_a')}\, .
\ee
\end{widetext}
Next, we notice that the integrand is invariant under translations $x_a\to x_a+\text{const.}$ and similarly under translations $x'_a\to x_a'+\text{const.}$, so we get a factor $L$ for each of the two translational symmetries and we can fix $x_1=x_1'=0$ in the integrand. Furthermore, we take the thermodynamic limit $\sum_\lambda\to \frac{L}{2\pi} \int \dd \lambda$ and notice that the integrand is invariant under global rapidity shifts $\lambda_a\to \lambda_a+\text{const.}$, hence it is convenient to change variables as $\chi_1=-N^{-1}\sum_a\lambda_a$ and $\chi_{a>1}=\lambda_a-N^{-1}\sum_a\lambda_a$. The change of coordinate $\sum_{a'}\mathcal{M}_{a,a'}\chi_{a'}=\lambda_a$ has a non-trivial Jacobian which must be taken into account when changing variables. In particular, $\det \mathcal{M}=-N^{-1}$. 

The integrand does not depend on $\chi_1$, hence one can explicitly integrate over $\chi_1$, getting a $\dd\lambda$ overall factor, i.e. the length of the interval on which we are averaging. Lastly, we change variables rescaling $y_a=|c| x_a$, $y_a'=|c|x_a'$ and $\mu_a=\chi_a/|c|$. Notice that, since $\lambda_a$ lived in an interval of width $\dd\lambda$, $\mu_a$ belongs on an interval of length $\dd\lambda/|c|$, which diverges with $|c|\to 0$.
Hence, in the $|c|\to 0$ limit one gets
\begin{widetext}
\be
\bar{P}=\left(\frac{\dd\lambda L}{2\pi}\right)^{-N} \frac{L\dd\lambda N!}{(2\pi)^N}N\int_{-\infty}^{+\infty} \dd^{N-1}\mu
\int_{-\infty}^{+\infty} \dd^{N-1} y \int_{-\infty}^{+\infty} \dd^{N-1} y'\, \Phi(0,y_2,...,y_N)\Phi(0,y_2',...,y'_N) e^{i\sum_{a=2}^N \mu_a (y_a-y_a')}\, .
\ee
\end{widetext}
Now, we could first integrate in the coordinates $y_a$ $y_a'$ and then in the the variables $\mu_a$. If one proceeds in this way, a decaying function of the $\mu_a$ variables is found. In terms of the original rapidities $\lambda_a$, this means the function decays as $\lambda_a$ are dragged apart from their center of mass on a typical length scale $\sim|c|$. In other words, the function is very peaked in the $\lambda-$space: this will be used in the forthcoming section. However, for the time being it is better to integrate first in the $\mu_a$ coordinates: this results in $N-1$ Dirac deltas that enforce $y_a=y'_a$
\be
\bar{P}=\left(\frac{\dd\lambda L}{2\pi}\right)^{-N} N\frac{L\dd\lambda N!}{2\pi}
\int_{-\infty}^{+\infty} \dd^{N-1} y \, |\Phi(0,y_2,...,y_N)|^2\, .
\ee
Finally, one notices that $\int_{-\infty}^{+\infty} \dd^{N-1} y \, |\Phi(0,y_2,...,y_N)|^2=1$ because of the normalization of the bound state wavefunction.
The final simple result is Eq. \eqref{eq_24}.

\bibliography{biblio}

\begin{thebibliography}{97}%
\makeatletter
\providecommand \@ifxundefined [1]{%
 \@ifx{#1\undefined}
}%
\providecommand \@ifnum [1]{%
 \ifnum #1\expandafter \@firstoftwo
 \else \expandafter \@secondoftwo
 \fi
}%
\providecommand \@ifx [1]{%
 \ifx #1\expandafter \@firstoftwo
 \else \expandafter \@secondoftwo
 \fi
}%
\providecommand \natexlab [1]{#1}%
\providecommand \enquote  [1]{``#1''}%
\providecommand \bibnamefont  [1]{#1}%
\providecommand \bibfnamefont [1]{#1}%
\providecommand \citenamefont [1]{#1}%
\providecommand \href@noop [0]{\@secondoftwo}%
\providecommand \href [0]{\begingroup \@sanitize@url \@href}%
\providecommand \@href[1]{\@@startlink{#1}\@@href}%
\providecommand \@@href[1]{\endgroup#1\@@endlink}%
\providecommand \@sanitize@url [0]{\catcode `\\12\catcode `\$12\catcode
  `\&12\catcode `\#12\catcode `\^12\catcode `\_12\catcode `\%12\relax}%
\providecommand \@@startlink[1]{}%
\providecommand \@@endlink[0]{}%
\providecommand \url  [0]{\begingroup\@sanitize@url \@url }%
\providecommand \@url [1]{\endgroup\@href {#1}{\urlprefix }}%
\providecommand \urlprefix  [0]{URL }%
\providecommand \Eprint [0]{\href }%
\providecommand \doibase [0]{http://dx.doi.org/}%
\providecommand \selectlanguage [0]{\@gobble}%
\providecommand \bibinfo  [0]{\@secondoftwo}%
\providecommand \bibfield  [0]{\@secondoftwo}%
\providecommand \translation [1]{[#1]}%
\providecommand \BibitemOpen [0]{}%
\providecommand \bibitemStop [0]{}%
\providecommand \bibitemNoStop [0]{.\EOS\space}%
\providecommand \EOS [0]{\spacefactor3000\relax}%
\providecommand \BibitemShut  [1]{\csname bibitem#1\endcsname}%
\let\auto@bib@innerbib\@empty
\bibitem [{\citenamefont {Bloch}(2005)}]{Bloch2005}%
  \BibitemOpen
  \bibfield  {author} {\bibinfo {author} {\bibfnamefont {I.}~\bibnamefont
  {Bloch}},\ }\href {\doibase 10.1038/nphys138} {\bibfield  {journal} {\bibinfo
   {journal} {Nature Physics}\ }\textbf {\bibinfo {volume} {1}},\ \bibinfo
  {pages} {23} (\bibinfo {year} {2005})}\BibitemShut {NoStop}%
\bibitem [{\citenamefont {Gring}\ \emph {et~al.}(2012)\citenamefont {Gring},
  \citenamefont {Kuhnert}, \citenamefont {Langen}, \citenamefont {Kitagawa},
  \citenamefont {Rauer}, \citenamefont {Schreitl}, \citenamefont {Mazets},
  \citenamefont {Smith}, \citenamefont {Demler},\ and\ \citenamefont
  {Schmiedmayer}}]{Gring1318}%
  \BibitemOpen
  \bibfield  {author} {\bibinfo {author} {\bibfnamefont {M.}~\bibnamefont
  {Gring}}, \bibinfo {author} {\bibfnamefont {M.}~\bibnamefont {Kuhnert}},
  \bibinfo {author} {\bibfnamefont {T.}~\bibnamefont {Langen}}, \bibinfo
  {author} {\bibfnamefont {T.}~\bibnamefont {Kitagawa}}, \bibinfo {author}
  {\bibfnamefont {B.}~\bibnamefont {Rauer}}, \bibinfo {author} {\bibfnamefont
  {M.}~\bibnamefont {Schreitl}}, \bibinfo {author} {\bibfnamefont
  {I.}~\bibnamefont {Mazets}}, \bibinfo {author} {\bibfnamefont {D.~A.}\
  \bibnamefont {Smith}}, \bibinfo {author} {\bibfnamefont {E.}~\bibnamefont
  {Demler}}, \ and\ \bibinfo {author} {\bibfnamefont {J.}~\bibnamefont
  {Schmiedmayer}},\ }\href {\doibase 10.1126/science.1224953} {\bibfield
  {journal} {\bibinfo  {journal} {Science}\ }\textbf {\bibinfo {volume}
  {337}},\ \bibinfo {pages} {1318} (\bibinfo {year} {2012})}\BibitemShut
  {NoStop}%
\bibitem [{\citenamefont {Schneider}\ \emph {et~al.}(2012)\citenamefont
  {Schneider}, \citenamefont {Hackerm{\"u}ller}, \citenamefont {Ronzheimer},
  \citenamefont {Will}, \citenamefont {Braun}, \citenamefont {Best},
  \citenamefont {Bloch}, \citenamefont {Demler}, \citenamefont {Mandt},
  \citenamefont {Rasch},\ and\ \citenamefont {Rosch}}]{Schneider2012}%
  \BibitemOpen
  \bibfield  {author} {\bibinfo {author} {\bibfnamefont {U.}~\bibnamefont
  {Schneider}}, \bibinfo {author} {\bibfnamefont {L.}~\bibnamefont
  {Hackerm{\"u}ller}}, \bibinfo {author} {\bibfnamefont {J.~P.}\ \bibnamefont
  {Ronzheimer}}, \bibinfo {author} {\bibfnamefont {S.}~\bibnamefont {Will}},
  \bibinfo {author} {\bibfnamefont {S.}~\bibnamefont {Braun}}, \bibinfo
  {author} {\bibfnamefont {T.}~\bibnamefont {Best}}, \bibinfo {author}
  {\bibfnamefont {I.}~\bibnamefont {Bloch}}, \bibinfo {author} {\bibfnamefont
  {E.}~\bibnamefont {Demler}}, \bibinfo {author} {\bibfnamefont
  {S.}~\bibnamefont {Mandt}}, \bibinfo {author} {\bibfnamefont
  {D.}~\bibnamefont {Rasch}}, \ and\ \bibinfo {author} {\bibfnamefont
  {A.}~\bibnamefont {Rosch}},\ }\href {\doibase 10.1038/nphys2205} {\bibfield
  {journal} {\bibinfo  {journal} {Nature Physics}\ }\textbf {\bibinfo {volume}
  {8}},\ \bibinfo {pages} {213} (\bibinfo {year} {2012})}\BibitemShut {NoStop}%
\bibitem [{\citenamefont {Langen}\ \emph {et~al.}(2015)\citenamefont {Langen},
  \citenamefont {Erne}, \citenamefont {Geiger}, \citenamefont {Rauer},
  \citenamefont {Schweigler}, \citenamefont {Kuhnert}, \citenamefont
  {Rohringer}, \citenamefont {Mazets}, \citenamefont {Gasenzer},\ and\
  \citenamefont {Schmiedmayer}}]{Langen207}%
  \BibitemOpen
  \bibfield  {author} {\bibinfo {author} {\bibfnamefont {T.}~\bibnamefont
  {Langen}}, \bibinfo {author} {\bibfnamefont {S.}~\bibnamefont {Erne}},
  \bibinfo {author} {\bibfnamefont {R.}~\bibnamefont {Geiger}}, \bibinfo
  {author} {\bibfnamefont {B.}~\bibnamefont {Rauer}}, \bibinfo {author}
  {\bibfnamefont {T.}~\bibnamefont {Schweigler}}, \bibinfo {author}
  {\bibfnamefont {M.}~\bibnamefont {Kuhnert}}, \bibinfo {author} {\bibfnamefont
  {W.}~\bibnamefont {Rohringer}}, \bibinfo {author} {\bibfnamefont {I.~E.}\
  \bibnamefont {Mazets}}, \bibinfo {author} {\bibfnamefont {T.}~\bibnamefont
  {Gasenzer}}, \ and\ \bibinfo {author} {\bibfnamefont {J.}~\bibnamefont
  {Schmiedmayer}},\ }\href {\doibase 10.1126/science.1257026} {\bibfield
  {journal} {\bibinfo  {journal} {Science}\ }\textbf {\bibinfo {volume}
  {348}},\ \bibinfo {pages} {207} (\bibinfo {year} {2015})}\BibitemShut
  {NoStop}%
\bibitem [{\citenamefont {Fukuhara}\ \emph
  {et~al.}(2013{\natexlab{a}})\citenamefont {Fukuhara}, \citenamefont
  {Kantian}, \citenamefont {Endres}, \citenamefont {Cheneau}, \citenamefont
  {Schau{\ss}}, \citenamefont {Hild}, \citenamefont {Bellem}, \citenamefont
  {Schollw{\"o}ck}, \citenamefont {Giamarchi}, \citenamefont {Gross},
  \citenamefont {Bloch},\ and\ \citenamefont {Kuhr}}]{Fukuhara2013}%
  \BibitemOpen
  \bibfield  {author} {\bibinfo {author} {\bibfnamefont {T.}~\bibnamefont
  {Fukuhara}}, \bibinfo {author} {\bibfnamefont {A.}~\bibnamefont {Kantian}},
  \bibinfo {author} {\bibfnamefont {M.}~\bibnamefont {Endres}}, \bibinfo
  {author} {\bibfnamefont {M.}~\bibnamefont {Cheneau}}, \bibinfo {author}
  {\bibfnamefont {P.}~\bibnamefont {Schau{\ss}}}, \bibinfo {author}
  {\bibfnamefont {S.}~\bibnamefont {Hild}}, \bibinfo {author} {\bibfnamefont
  {D.}~\bibnamefont {Bellem}}, \bibinfo {author} {\bibfnamefont
  {U.}~\bibnamefont {Schollw{\"o}ck}}, \bibinfo {author} {\bibfnamefont
  {T.}~\bibnamefont {Giamarchi}}, \bibinfo {author} {\bibfnamefont
  {C.}~\bibnamefont {Gross}}, \bibinfo {author} {\bibfnamefont
  {I.}~\bibnamefont {Bloch}}, \ and\ \bibinfo {author} {\bibfnamefont
  {S.}~\bibnamefont {Kuhr}},\ }\href {\doibase 10.1038/nphys2561} {\bibfield
  {journal} {\bibinfo  {journal} {Nature Physics}\ }\textbf {\bibinfo {volume}
  {9}},\ \bibinfo {pages} {235} (\bibinfo {year}
  {2013}{\natexlab{a}})}\BibitemShut {NoStop}%
\bibitem [{\citenamefont {Jepsen}\ \emph {et~al.}(2020)\citenamefont {Jepsen},
  \citenamefont {Amato-Grill}, \citenamefont {Dimitrova}, \citenamefont {Ho},
  \citenamefont {Demler},\ and\ \citenamefont {Ketterle}}]{Jepsen2020}%
  \BibitemOpen
  \bibfield  {author} {\bibinfo {author} {\bibfnamefont {P.~N.}\ \bibnamefont
  {Jepsen}}, \bibinfo {author} {\bibfnamefont {J.}~\bibnamefont {Amato-Grill}},
  \bibinfo {author} {\bibfnamefont {I.}~\bibnamefont {Dimitrova}}, \bibinfo
  {author} {\bibfnamefont {W.~W.}\ \bibnamefont {Ho}}, \bibinfo {author}
  {\bibfnamefont {E.}~\bibnamefont {Demler}}, \ and\ \bibinfo {author}
  {\bibfnamefont {W.}~\bibnamefont {Ketterle}},\ }\href {\doibase
  10.1038/s41586-020-3033-y} {\bibfield  {journal} {\bibinfo  {journal}
  {Nature}\ }\textbf {\bibinfo {volume} {588}},\ \bibinfo {pages} {403}
  (\bibinfo {year} {2020})}\BibitemShut {NoStop}%
\bibitem [{\citenamefont {Hild}\ \emph {et~al.}(2014)\citenamefont {Hild},
  \citenamefont {Fukuhara}, \citenamefont {Schau\ss{}}, \citenamefont {Zeiher},
  \citenamefont {Knap}, \citenamefont {Demler}, \citenamefont {Bloch},\ and\
  \citenamefont {Gross}}]{PhysRevLett.113.147205}%
  \BibitemOpen
  \bibfield  {author} {\bibinfo {author} {\bibfnamefont {S.}~\bibnamefont
  {Hild}}, \bibinfo {author} {\bibfnamefont {T.}~\bibnamefont {Fukuhara}},
  \bibinfo {author} {\bibfnamefont {P.}~\bibnamefont {Schau\ss{}}}, \bibinfo
  {author} {\bibfnamefont {J.}~\bibnamefont {Zeiher}}, \bibinfo {author}
  {\bibfnamefont {M.}~\bibnamefont {Knap}}, \bibinfo {author} {\bibfnamefont
  {E.}~\bibnamefont {Demler}}, \bibinfo {author} {\bibfnamefont
  {I.}~\bibnamefont {Bloch}}, \ and\ \bibinfo {author} {\bibfnamefont
  {C.}~\bibnamefont {Gross}},\ }\href {\doibase 10.1103/PhysRevLett.113.147205}
  {\bibfield  {journal} {\bibinfo  {journal} {Phys. Rev. Lett.}\ }\textbf
  {\bibinfo {volume} {113}},\ \bibinfo {pages} {147205} (\bibinfo {year}
  {2014})}\BibitemShut {NoStop}%
\bibitem [{\citenamefont {Fukuhara}\ \emph
  {et~al.}(2013{\natexlab{b}})\citenamefont {Fukuhara}, \citenamefont
  {Schau{\ss}}, \citenamefont {Endres}, \citenamefont {Hild}, \citenamefont
  {Cheneau}, \citenamefont {Bloch},\ and\ \citenamefont
  {Gross}}]{Fukuhara2013_bis}%
  \BibitemOpen
  \bibfield  {author} {\bibinfo {author} {\bibfnamefont {T.}~\bibnamefont
  {Fukuhara}}, \bibinfo {author} {\bibfnamefont {P.}~\bibnamefont
  {Schau{\ss}}}, \bibinfo {author} {\bibfnamefont {M.}~\bibnamefont {Endres}},
  \bibinfo {author} {\bibfnamefont {S.}~\bibnamefont {Hild}}, \bibinfo {author}
  {\bibfnamefont {M.}~\bibnamefont {Cheneau}}, \bibinfo {author} {\bibfnamefont
  {I.}~\bibnamefont {Bloch}}, \ and\ \bibinfo {author} {\bibfnamefont
  {C.}~\bibnamefont {Gross}},\ }\href {\doibase 10.1038/nature12541} {\bibfield
   {journal} {\bibinfo  {journal} {Nature}\ }\textbf {\bibinfo {volume}
  {502}},\ \bibinfo {pages} {76} (\bibinfo {year}
  {2013}{\natexlab{b}})}\BibitemShut {NoStop}%
\bibitem [{\citenamefont {Lieb}\ and\ \citenamefont
  {Liniger}(1963)}]{PhysRev.130.1605}%
  \BibitemOpen
  \bibfield  {author} {\bibinfo {author} {\bibfnamefont {E.~H.}\ \bibnamefont
  {Lieb}}\ and\ \bibinfo {author} {\bibfnamefont {W.}~\bibnamefont {Liniger}},\
  }\href {\doibase 10.1103/PhysRev.130.1605} {\bibfield  {journal} {\bibinfo
  {journal} {Phys. Rev.}\ }\textbf {\bibinfo {volume} {130}},\ \bibinfo {pages}
  {1605} (\bibinfo {year} {1963})}\BibitemShut {NoStop}%
\bibitem [{\citenamefont {Lieb}(1963)}]{PhysRev.130.1616}%
  \BibitemOpen
  \bibfield  {author} {\bibinfo {author} {\bibfnamefont {E.~H.}\ \bibnamefont
  {Lieb}},\ }\href {\doibase 10.1103/PhysRev.130.1616} {\bibfield  {journal}
  {\bibinfo  {journal} {Phys. Rev.}\ }\textbf {\bibinfo {volume} {130}},\
  \bibinfo {pages} {1616} (\bibinfo {year} {1963})}\BibitemShut {NoStop}%
\bibitem [{\citenamefont {McGuire}(1964)}]{doi:10.1063/1.1704156}%
  \BibitemOpen
  \bibfield  {author} {\bibinfo {author} {\bibfnamefont {J.~B.}\ \bibnamefont
  {McGuire}},\ }\href {\doibase 10.1063/1.1704156} {\bibfield  {journal}
  {\bibinfo  {journal} {Journal of Mathematical Physics}\ }\textbf {\bibinfo
  {volume} {5}},\ \bibinfo {pages} {622} (\bibinfo {year} {1964})},\ \Eprint
  {http://arxiv.org/abs/https://doi.org/10.1063/1.1704156}
  {https://doi.org/10.1063/1.1704156} \BibitemShut {NoStop}%
\bibitem [{\citenamefont {Olshanii}(1998)}]{PhysRevLett.81.938}%
  \BibitemOpen
  \bibfield  {author} {\bibinfo {author} {\bibfnamefont {M.}~\bibnamefont
  {Olshanii}},\ }\href {\doibase 10.1103/PhysRevLett.81.938} {\bibfield
  {journal} {\bibinfo  {journal} {Phys. Rev. Lett.}\ }\textbf {\bibinfo
  {volume} {81}},\ \bibinfo {pages} {938} (\bibinfo {year} {1998})}\BibitemShut
  {NoStop}%
\bibitem [{\citenamefont {Bloch}\ \emph {et~al.}(2008)\citenamefont {Bloch},
  \citenamefont {Dalibard},\ and\ \citenamefont {Zwerger}}]{RevModPhys.80.885}%
  \BibitemOpen
  \bibfield  {author} {\bibinfo {author} {\bibfnamefont {I.}~\bibnamefont
  {Bloch}}, \bibinfo {author} {\bibfnamefont {J.}~\bibnamefont {Dalibard}}, \
  and\ \bibinfo {author} {\bibfnamefont {W.}~\bibnamefont {Zwerger}},\ }\href
  {\doibase 10.1103/RevModPhys.80.885} {\bibfield  {journal} {\bibinfo
  {journal} {Rev. Mod. Phys.}\ }\textbf {\bibinfo {volume} {80}},\ \bibinfo
  {pages} {885} (\bibinfo {year} {2008})}\BibitemShut {NoStop}%
\bibitem [{\citenamefont {Kinoshita}\ \emph {et~al.}(2004)\citenamefont
  {Kinoshita}, \citenamefont {Wenger},\ and\ \citenamefont
  {Weiss}}]{Kinoshita1125}%
  \BibitemOpen
  \bibfield  {author} {\bibinfo {author} {\bibfnamefont {T.}~\bibnamefont
  {Kinoshita}}, \bibinfo {author} {\bibfnamefont {T.}~\bibnamefont {Wenger}}, \
  and\ \bibinfo {author} {\bibfnamefont {D.~S.}\ \bibnamefont {Weiss}},\ }\href
  {\doibase 10.1126/science.1100700} {\bibfield  {journal} {\bibinfo  {journal}
  {Science}\ }\textbf {\bibinfo {volume} {305}},\ \bibinfo {pages} {1125}
  (\bibinfo {year} {2004})}\BibitemShut {NoStop}%
\bibitem [{\citenamefont {Kinoshita}\ \emph {et~al.}(2005)\citenamefont
  {Kinoshita}, \citenamefont {Wenger},\ and\ \citenamefont
  {Weiss}}]{PhysRevLett.95.190406}%
  \BibitemOpen
  \bibfield  {author} {\bibinfo {author} {\bibfnamefont {T.}~\bibnamefont
  {Kinoshita}}, \bibinfo {author} {\bibfnamefont {T.}~\bibnamefont {Wenger}}, \
  and\ \bibinfo {author} {\bibfnamefont {D.~S.}\ \bibnamefont {Weiss}},\ }\href
  {\doibase 10.1103/PhysRevLett.95.190406} {\bibfield  {journal} {\bibinfo
  {journal} {Phys. Rev. Lett.}\ }\textbf {\bibinfo {volume} {95}},\ \bibinfo
  {pages} {190406} (\bibinfo {year} {2005})}\BibitemShut {NoStop}%
\bibitem [{\citenamefont {Kinoshita}\ \emph {et~al.}(2006)\citenamefont
  {Kinoshita}, \citenamefont {Wenger},\ and\ \citenamefont
  {Weiss}}]{Kinoshita2006}%
  \BibitemOpen
  \bibfield  {author} {\bibinfo {author} {\bibfnamefont {T.}~\bibnamefont
  {Kinoshita}}, \bibinfo {author} {\bibfnamefont {T.}~\bibnamefont {Wenger}}, \
  and\ \bibinfo {author} {\bibfnamefont {D.~S.}\ \bibnamefont {Weiss}},\ }\href
  {\doibase 10.1038/nature04693} {\bibfield  {journal} {\bibinfo  {journal}
  {Nature}\ }\textbf {\bibinfo {volume} {440}},\ \bibinfo {pages} {900}
  (\bibinfo {year} {2006})}\BibitemShut {NoStop}%
\bibitem [{\citenamefont {Armijo}\ \emph {et~al.}(2010)\citenamefont {Armijo},
  \citenamefont {Jacqmin}, \citenamefont {Kheruntsyan},\ and\ \citenamefont
  {Bouchoule}}]{PhysRevLett.105.230402}%
  \BibitemOpen
  \bibfield  {author} {\bibinfo {author} {\bibfnamefont {J.}~\bibnamefont
  {Armijo}}, \bibinfo {author} {\bibfnamefont {T.}~\bibnamefont {Jacqmin}},
  \bibinfo {author} {\bibfnamefont {K.~V.}\ \bibnamefont {Kheruntsyan}}, \ and\
  \bibinfo {author} {\bibfnamefont {I.}~\bibnamefont {Bouchoule}},\ }\href
  {\doibase 10.1103/PhysRevLett.105.230402} {\bibfield  {journal} {\bibinfo
  {journal} {Phys. Rev. Lett.}\ }\textbf {\bibinfo {volume} {105}},\ \bibinfo
  {pages} {230402} (\bibinfo {year} {2010})}\BibitemShut {NoStop}%
\bibitem [{\citenamefont {van Amerongen}\ \emph {et~al.}(2008)\citenamefont
  {van Amerongen}, \citenamefont {van Es}, \citenamefont {Wicke}, \citenamefont
  {Kheruntsyan},\ and\ \citenamefont {van Druten}}]{PhysRevLett.100.090402}%
  \BibitemOpen
  \bibfield  {author} {\bibinfo {author} {\bibfnamefont {A.~H.}\ \bibnamefont
  {van Amerongen}}, \bibinfo {author} {\bibfnamefont {J.~J.~P.}\ \bibnamefont
  {van Es}}, \bibinfo {author} {\bibfnamefont {P.}~\bibnamefont {Wicke}},
  \bibinfo {author} {\bibfnamefont {K.~V.}\ \bibnamefont {Kheruntsyan}}, \ and\
  \bibinfo {author} {\bibfnamefont {N.~J.}\ \bibnamefont {van Druten}},\ }\href
  {\doibase 10.1103/PhysRevLett.100.090402} {\bibfield  {journal} {\bibinfo
  {journal} {Phys. Rev. Lett.}\ }\textbf {\bibinfo {volume} {100}},\ \bibinfo
  {pages} {090402} (\bibinfo {year} {2008})}\BibitemShut {NoStop}%
\bibitem [{\citenamefont {Schemmer}\ \emph {et~al.}(2019)\citenamefont
  {Schemmer}, \citenamefont {Bouchoule}, \citenamefont {Doyon},\ and\
  \citenamefont {Dubail}}]{PhysRevLett.122.090601}%
  \BibitemOpen
  \bibfield  {author} {\bibinfo {author} {\bibfnamefont {M.}~\bibnamefont
  {Schemmer}}, \bibinfo {author} {\bibfnamefont {I.}~\bibnamefont {Bouchoule}},
  \bibinfo {author} {\bibfnamefont {B.}~\bibnamefont {Doyon}}, \ and\ \bibinfo
  {author} {\bibfnamefont {J.}~\bibnamefont {Dubail}},\ }\href {\doibase
  10.1103/PhysRevLett.122.090601} {\bibfield  {journal} {\bibinfo  {journal}
  {Phys. Rev. Lett.}\ }\textbf {\bibinfo {volume} {122}},\ \bibinfo {pages}
  {090601} (\bibinfo {year} {2019})}\BibitemShut {NoStop}%
\bibitem [{\citenamefont {Haller}\ \emph {et~al.}(2011)\citenamefont {Haller},
  \citenamefont {Rabie}, \citenamefont {Mark}, \citenamefont {Danzl},
  \citenamefont {Hart}, \citenamefont {Lauber}, \citenamefont {Pupillo},\ and\
  \citenamefont {N\"agerl}}]{PhysRevLett.107.230404}%
  \BibitemOpen
  \bibfield  {author} {\bibinfo {author} {\bibfnamefont {E.}~\bibnamefont
  {Haller}}, \bibinfo {author} {\bibfnamefont {M.}~\bibnamefont {Rabie}},
  \bibinfo {author} {\bibfnamefont {M.~J.}\ \bibnamefont {Mark}}, \bibinfo
  {author} {\bibfnamefont {J.~G.}\ \bibnamefont {Danzl}}, \bibinfo {author}
  {\bibfnamefont {R.}~\bibnamefont {Hart}}, \bibinfo {author} {\bibfnamefont
  {K.}~\bibnamefont {Lauber}}, \bibinfo {author} {\bibfnamefont
  {G.}~\bibnamefont {Pupillo}}, \ and\ \bibinfo {author} {\bibfnamefont
  {H.-C.}\ \bibnamefont {N\"agerl}},\ }\href {\doibase
  10.1103/PhysRevLett.107.230404} {\bibfield  {journal} {\bibinfo  {journal}
  {Phys. Rev. Lett.}\ }\textbf {\bibinfo {volume} {107}},\ \bibinfo {pages}
  {230404} (\bibinfo {year} {2011})}\BibitemShut {NoStop}%
\bibitem [{\citenamefont {Malvania}\ \emph {et~al.}(2020)\citenamefont
  {Malvania}, \citenamefont {Zhang}, \citenamefont {Le}, \citenamefont
  {Dubail}, \citenamefont {Rigol},\ and\ \citenamefont
  {Weiss}}]{malvania2020generalized}%
  \BibitemOpen
  \bibfield  {author} {\bibinfo {author} {\bibfnamefont {N.}~\bibnamefont
  {Malvania}}, \bibinfo {author} {\bibfnamefont {Y.}~\bibnamefont {Zhang}},
  \bibinfo {author} {\bibfnamefont {Y.}~\bibnamefont {Le}}, \bibinfo {author}
  {\bibfnamefont {J.}~\bibnamefont {Dubail}}, \bibinfo {author} {\bibfnamefont
  {M.}~\bibnamefont {Rigol}}, \ and\ \bibinfo {author} {\bibfnamefont {D.~S.}\
  \bibnamefont {Weiss}},\ }\href@noop {} {} (\bibinfo {year} {2020}),\ \Eprint
  {http://arxiv.org/abs/2009.06651} {arXiv:2009.06651 [cond-mat.quant-gas]}
  \BibitemShut {NoStop}%
\bibitem [{\citenamefont {Haller}\ \emph {et~al.}(2009)\citenamefont {Haller},
  \citenamefont {Gustavsson}, \citenamefont {Mark}, \citenamefont {Danzl},
  \citenamefont {Hart}, \citenamefont {Pupillo},\ and\ \citenamefont
  {N{\"a}gerl}}]{Haller1224}%
  \BibitemOpen
  \bibfield  {author} {\bibinfo {author} {\bibfnamefont {E.}~\bibnamefont
  {Haller}}, \bibinfo {author} {\bibfnamefont {M.}~\bibnamefont {Gustavsson}},
  \bibinfo {author} {\bibfnamefont {M.~J.}\ \bibnamefont {Mark}}, \bibinfo
  {author} {\bibfnamefont {J.~G.}\ \bibnamefont {Danzl}}, \bibinfo {author}
  {\bibfnamefont {R.}~\bibnamefont {Hart}}, \bibinfo {author} {\bibfnamefont
  {G.}~\bibnamefont {Pupillo}}, \ and\ \bibinfo {author} {\bibfnamefont
  {H.-C.}\ \bibnamefont {N{\"a}gerl}},\ }\href {\doibase
  10.1126/science.1175850} {\bibfield  {journal} {\bibinfo  {journal}
  {Science}\ }\textbf {\bibinfo {volume} {325}},\ \bibinfo {pages} {1224}
  (\bibinfo {year} {2009})}\BibitemShut {NoStop}%
\bibitem [{\citenamefont {Solano}\ \emph {et~al.}(2019)\citenamefont {Solano},
  \citenamefont {Duan}, \citenamefont {Chen}, \citenamefont {Rudelis},
  \citenamefont {Chin},\ and\ \citenamefont {Vuleti\ifmmode~\acute{c}\else
  \'{c}\fi{}}}]{PabloSolanoExp2019}%
  \BibitemOpen
  \bibfield  {author} {\bibinfo {author} {\bibfnamefont {P.}~\bibnamefont
  {Solano}}, \bibinfo {author} {\bibfnamefont {Y.}~\bibnamefont {Duan}},
  \bibinfo {author} {\bibfnamefont {Y.-T.}\ \bibnamefont {Chen}}, \bibinfo
  {author} {\bibfnamefont {A.}~\bibnamefont {Rudelis}}, \bibinfo {author}
  {\bibfnamefont {C.}~\bibnamefont {Chin}}, \ and\ \bibinfo {author}
  {\bibfnamefont {V.}~\bibnamefont {Vuleti\ifmmode~\acute{c}\else
  \'{c}\fi{}}},\ }\href {\doibase 10.1103/PhysRevLett.123.173401} {\bibfield
  {journal} {\bibinfo  {journal} {Phys. Rev. Lett.}\ }\textbf {\bibinfo
  {volume} {123}},\ \bibinfo {pages} {173401} (\bibinfo {year}
  {2019})}\BibitemShut {NoStop}%
\bibitem [{\citenamefont {Kao}\ \emph {et~al.}(2021)\citenamefont {Kao},
  \citenamefont {Li}, \citenamefont {Lin}, \citenamefont {Gopalakrishnan},\
  and\ \citenamefont {Lev}}]{kao2020creating}%
  \BibitemOpen
  \bibfield  {author} {\bibinfo {author} {\bibfnamefont {W.}~\bibnamefont
  {Kao}}, \bibinfo {author} {\bibfnamefont {K.-Y.}\ \bibnamefont {Li}},
  \bibinfo {author} {\bibfnamefont {K.-Y.}\ \bibnamefont {Lin}}, \bibinfo
  {author} {\bibfnamefont {S.}~\bibnamefont {Gopalakrishnan}}, \ and\ \bibinfo
  {author} {\bibfnamefont {B.~L.}\ \bibnamefont {Lev}},\ }\href {\doibase
  10.1126/science.abb4928} {\bibfield  {journal} {\bibinfo  {journal}
  {Science}\ }\textbf {\bibinfo {volume} {371}},\ \bibinfo {pages} {296}
  (\bibinfo {year} {2021})}\BibitemShut {NoStop}%
\bibitem [{\citenamefont {Korepin}\ \emph {et~al.}(1997)\citenamefont
  {Korepin}, \citenamefont {Bogoliubov},\ and\ \citenamefont
  {Izergin}}]{korepin1997quantum}%
  \BibitemOpen
  \bibfield  {author} {\bibinfo {author} {\bibfnamefont {V.~E.}\ \bibnamefont
  {Korepin}}, \bibinfo {author} {\bibfnamefont {N.~M.}\ \bibnamefont
  {Bogoliubov}}, \ and\ \bibinfo {author} {\bibfnamefont {A.~G.}\ \bibnamefont
  {Izergin}},\ }\href@noop {} {\emph {\bibinfo {title} {Quantum inverse
  scattering method and correlation functions}}},\ Vol.~\bibinfo {volume} {3}\
  (\bibinfo  {publisher} {Cambridge university press},\ \bibinfo {year}
  {1997})\BibitemShut {NoStop}%
\bibitem [{\citenamefont {Mussardo}(2010)}]{Mussardo:2010mgq}%
  \BibitemOpen
  \bibfield  {author} {\bibinfo {author} {\bibfnamefont {G.}~\bibnamefont
  {Mussardo}},\ }\href@noop {} {\emph {\bibinfo {title} {{Statistical field
  theory}}}}\ (\bibinfo  {publisher} {Oxford Univ. Press},\ \bibinfo {address}
  {New York, NY},\ \bibinfo {year} {2010})\BibitemShut {NoStop}%
\bibitem [{\citenamefont {Polkovnikov}\ \emph {et~al.}(2011)\citenamefont
  {Polkovnikov}, \citenamefont {Sengupta}, \citenamefont {Silva},\ and\
  \citenamefont {Vengalattore}}]{RevModPhys.83.863}%
  \BibitemOpen
  \bibfield  {author} {\bibinfo {author} {\bibfnamefont {A.}~\bibnamefont
  {Polkovnikov}}, \bibinfo {author} {\bibfnamefont {K.}~\bibnamefont
  {Sengupta}}, \bibinfo {author} {\bibfnamefont {A.}~\bibnamefont {Silva}}, \
  and\ \bibinfo {author} {\bibfnamefont {M.}~\bibnamefont {Vengalattore}},\
  }\href {\doibase 10.1103/RevModPhys.83.863} {\bibfield  {journal} {\bibinfo
  {journal} {Rev. Mod. Phys.}\ }\textbf {\bibinfo {volume} {83}},\ \bibinfo
  {pages} {863} (\bibinfo {year} {2011})}\BibitemShut {NoStop}%
\bibitem [{\citenamefont {Calabrese}\ \emph {et~al.}(2016)\citenamefont
  {Calabrese}, \citenamefont {Essler},\ and\ \citenamefont
  {Mussardo}}]{Calabrese_2016}%
  \BibitemOpen
  \bibfield  {author} {\bibinfo {author} {\bibfnamefont {P.}~\bibnamefont
  {Calabrese}}, \bibinfo {author} {\bibfnamefont {F.~H.~L.}\ \bibnamefont
  {Essler}}, \ and\ \bibinfo {author} {\bibfnamefont {G.}~\bibnamefont
  {Mussardo}},\ }\href {\doibase 10.1088/1742-5468/2016/06/064001} {\bibfield
  {journal} {\bibinfo  {journal} {Journal of Statistical Mechanics: Theory and
  Experiment}\ }\textbf {\bibinfo {volume} {2016}},\ \bibinfo {pages} {064001}
  (\bibinfo {year} {2016})}\BibitemShut {NoStop}%
\bibitem [{\citenamefont {Bertini}\ \emph {et~al.}(2020)\citenamefont
  {Bertini}, \citenamefont {Heidrich-Meisner}, \citenamefont {Karrasch},
  \citenamefont {Prosen}, \citenamefont {Steinigeweg},\ and\ \citenamefont
  {Znidaric}}]{bertini2020finitetemperature}%
  \BibitemOpen
  \bibfield  {author} {\bibinfo {author} {\bibfnamefont {B.}~\bibnamefont
  {Bertini}}, \bibinfo {author} {\bibfnamefont {F.}~\bibnamefont
  {Heidrich-Meisner}}, \bibinfo {author} {\bibfnamefont {C.}~\bibnamefont
  {Karrasch}}, \bibinfo {author} {\bibfnamefont {T.}~\bibnamefont {Prosen}},
  \bibinfo {author} {\bibfnamefont {R.}~\bibnamefont {Steinigeweg}}, \ and\
  \bibinfo {author} {\bibfnamefont {M.}~\bibnamefont {Znidaric}},\ }\href@noop
  {} {} (\bibinfo {year} {2020}),\ \Eprint {http://arxiv.org/abs/2003.03334}
  {arXiv:2003.03334 [cond-mat.str-el]} \BibitemShut {NoStop}%
\bibitem [{\citenamefont {Rigol}\ \emph {et~al.}(2007)\citenamefont {Rigol},
  \citenamefont {Dunjko}, \citenamefont {Yurovsky},\ and\ \citenamefont
  {Olshanii}}]{rigol2007relaxation}%
  \BibitemOpen
  \bibfield  {author} {\bibinfo {author} {\bibfnamefont {M.}~\bibnamefont
  {Rigol}}, \bibinfo {author} {\bibfnamefont {V.}~\bibnamefont {Dunjko}},
  \bibinfo {author} {\bibfnamefont {V.}~\bibnamefont {Yurovsky}}, \ and\
  \bibinfo {author} {\bibfnamefont {M.}~\bibnamefont {Olshanii}},\ }\href
  {\doibase 10.1103/PhysRevLett.98.050405} {\bibfield  {journal} {\bibinfo
  {journal} {Phys. Rev. Lett.}\ }\textbf {\bibinfo {volume} {98}},\ \bibinfo
  {pages} {050405} (\bibinfo {year} {2007})}\BibitemShut {NoStop}%
\bibitem [{\citenamefont {Inouye}\ \emph {et~al.}(1998)\citenamefont {Inouye},
  \citenamefont {Andrews}, \citenamefont {Stenger}, \citenamefont {Miesner},
  \citenamefont {Stamper-Kurn},\ and\ \citenamefont {Ketterle}}]{Inouye1998}%
  \BibitemOpen
  \bibfield  {author} {\bibinfo {author} {\bibfnamefont {S.}~\bibnamefont
  {Inouye}}, \bibinfo {author} {\bibfnamefont {M.~R.}\ \bibnamefont {Andrews}},
  \bibinfo {author} {\bibfnamefont {J.}~\bibnamefont {Stenger}}, \bibinfo
  {author} {\bibfnamefont {H.-J.}\ \bibnamefont {Miesner}}, \bibinfo {author}
  {\bibfnamefont {D.~M.}\ \bibnamefont {Stamper-Kurn}}, \ and\ \bibinfo
  {author} {\bibfnamefont {W.}~\bibnamefont {Ketterle}},\ }\href {\doibase
  10.1038/32354} {\bibfield  {journal} {\bibinfo  {journal} {Nature}\ }\textbf
  {\bibinfo {volume} {392}},\ \bibinfo {pages} {151} (\bibinfo {year}
  {1998})}\BibitemShut {NoStop}%
\bibitem [{\citenamefont {Bergeman}\ \emph {et~al.}(2003)\citenamefont
  {Bergeman}, \citenamefont {Moore},\ and\ \citenamefont
  {Olshanii}}]{PhysRevLett.91.163201}%
  \BibitemOpen
  \bibfield  {author} {\bibinfo {author} {\bibfnamefont {T.}~\bibnamefont
  {Bergeman}}, \bibinfo {author} {\bibfnamefont {M.~G.}\ \bibnamefont {Moore}},
  \ and\ \bibinfo {author} {\bibfnamefont {M.}~\bibnamefont {Olshanii}},\
  }\href {\doibase 10.1103/PhysRevLett.91.163201} {\bibfield  {journal}
  {\bibinfo  {journal} {Phys. Rev. Lett.}\ }\textbf {\bibinfo {volume} {91}},\
  \bibinfo {pages} {163201} (\bibinfo {year} {2003})}\BibitemShut {NoStop}%
\bibitem [{\citenamefont {Takahashi}(2005)}]{takahashi2005thermodynamics}%
  \BibitemOpen
  \bibfield  {author} {\bibinfo {author} {\bibfnamefont {M.}~\bibnamefont
  {Takahashi}},\ }\href@noop {} {\emph {\bibinfo {title} {Thermodynamics of
  one-dimensional solvable models}}}\ (\bibinfo  {publisher} {Cambridge
  University Press},\ \bibinfo {year} {2005})\BibitemShut {NoStop}%
\bibitem [{\citenamefont {Calabrese}\ and\ \citenamefont
  {Caux}(2007{\natexlab{a}})}]{PhysRevLett.98.150403}%
  \BibitemOpen
  \bibfield  {author} {\bibinfo {author} {\bibfnamefont {P.}~\bibnamefont
  {Calabrese}}\ and\ \bibinfo {author} {\bibfnamefont {J.-S.}\ \bibnamefont
  {Caux}},\ }\href {\doibase 10.1103/PhysRevLett.98.150403} {\bibfield
  {journal} {\bibinfo  {journal} {Phys. Rev. Lett.}\ }\textbf {\bibinfo
  {volume} {98}},\ \bibinfo {pages} {150403} (\bibinfo {year}
  {2007}{\natexlab{a}})}\BibitemShut {NoStop}%
\bibitem [{\citenamefont {Calabrese}\ and\ \citenamefont
  {Caux}(2007{\natexlab{b}})}]{Calabrese_2007}%
  \BibitemOpen
  \bibfield  {author} {\bibinfo {author} {\bibfnamefont {P.}~\bibnamefont
  {Calabrese}}\ and\ \bibinfo {author} {\bibfnamefont {J.-S.}\ \bibnamefont
  {Caux}},\ }\href {\doibase 10.1088/1742-5468/2007/08/p08032} {\bibfield
  {journal} {\bibinfo  {journal} {Journal of Statistical Mechanics: Theory and
  Experiment}\ }\textbf {\bibinfo {volume} {2007}},\ \bibinfo {pages} {P08032}
  (\bibinfo {year} {2007}{\natexlab{b}})}\BibitemShut {NoStop}%
\bibitem [{\citenamefont {Sakmann}\ \emph {et~al.}(2005)\citenamefont
  {Sakmann}, \citenamefont {Streltsov}, \citenamefont {Alon},\ and\
  \citenamefont {Cederbaum}}]{PhysRevA.72.033613}%
  \BibitemOpen
  \bibfield  {author} {\bibinfo {author} {\bibfnamefont {K.}~\bibnamefont
  {Sakmann}}, \bibinfo {author} {\bibfnamefont {A.~I.}\ \bibnamefont
  {Streltsov}}, \bibinfo {author} {\bibfnamefont {O.~E.}\ \bibnamefont {Alon}},
  \ and\ \bibinfo {author} {\bibfnamefont {L.~S.}\ \bibnamefont {Cederbaum}},\
  }\href {\doibase 10.1103/PhysRevA.72.033613} {\bibfield  {journal} {\bibinfo
  {journal} {Phys. Rev. A}\ }\textbf {\bibinfo {volume} {72}},\ \bibinfo
  {pages} {033613} (\bibinfo {year} {2005})}\BibitemShut {NoStop}%
\bibitem [{\citenamefont {Calogero}\ and\ \citenamefont
  {Degasperis}(1975)}]{PhysRevA.11.265}%
  \BibitemOpen
  \bibfield  {author} {\bibinfo {author} {\bibfnamefont {F.}~\bibnamefont
  {Calogero}}\ and\ \bibinfo {author} {\bibfnamefont {A.}~\bibnamefont
  {Degasperis}},\ }\href {\doibase 10.1103/PhysRevA.11.265} {\bibfield
  {journal} {\bibinfo  {journal} {Phys. Rev. A}\ }\textbf {\bibinfo {volume}
  {11}},\ \bibinfo {pages} {265} (\bibinfo {year} {1975})}\BibitemShut
  {NoStop}%
\bibitem [{\citenamefont {Kanamoto}\ \emph {et~al.}(2006)\citenamefont
  {Kanamoto}, \citenamefont {Saito},\ and\ \citenamefont
  {Ueda}}]{PhysRevA.73.033611}%
  \BibitemOpen
  \bibfield  {author} {\bibinfo {author} {\bibfnamefont {R.}~\bibnamefont
  {Kanamoto}}, \bibinfo {author} {\bibfnamefont {H.}~\bibnamefont {Saito}}, \
  and\ \bibinfo {author} {\bibfnamefont {M.}~\bibnamefont {Ueda}},\ }\href
  {\doibase 10.1103/PhysRevA.73.033611} {\bibfield  {journal} {\bibinfo
  {journal} {Phys. Rev. A}\ }\textbf {\bibinfo {volume} {73}},\ \bibinfo
  {pages} {033611} (\bibinfo {year} {2006})}\BibitemShut {NoStop}%
\bibitem [{\citenamefont {Castin}\ and\ \citenamefont
  {Herzog}(2001)}]{CASTIN2001419}%
  \BibitemOpen
  \bibfield  {author} {\bibinfo {author} {\bibfnamefont {Y.}~\bibnamefont
  {Castin}}\ and\ \bibinfo {author} {\bibfnamefont {C.}~\bibnamefont
  {Herzog}},\ }\href {\doibase https://doi.org/10.1016/S1296-2147(01)01183-0}
  {\bibfield  {journal} {\bibinfo  {journal} {Comptes Rendus de l'Académie des
  Sciences - Series IV - Physics}\ }\textbf {\bibinfo {volume} {2}},\ \bibinfo
  {pages} {419 } (\bibinfo {year} {2001})}\BibitemShut {NoStop}%
\bibitem [{\citenamefont {Tempfli}\ \emph {et~al.}(2008)\citenamefont
  {Tempfli}, \citenamefont {Zöllner},\ and\ \citenamefont
  {Schmelcher}}]{Tempfli_2008}%
  \BibitemOpen
  \bibfield  {author} {\bibinfo {author} {\bibfnamefont {E.}~\bibnamefont
  {Tempfli}}, \bibinfo {author} {\bibfnamefont {S.}~\bibnamefont {Zöllner}}, \
  and\ \bibinfo {author} {\bibfnamefont {P.}~\bibnamefont {Schmelcher}},\
  }\href {\doibase 10.1088/1367-2630/10/10/103021} {\bibfield  {journal}
  {\bibinfo  {journal} {New Journal of Physics}\ }\textbf {\bibinfo {volume}
  {10}},\ \bibinfo {pages} {103021} (\bibinfo {year} {2008})}\BibitemShut
  {NoStop}%
\bibitem [{\citenamefont {Piroli}\ \emph
  {et~al.}(2016{\natexlab{a}})\citenamefont {Piroli}, \citenamefont
  {Calabrese},\ and\ \citenamefont {Essler}}]{PhysRevLett.116.070408}%
  \BibitemOpen
  \bibfield  {author} {\bibinfo {author} {\bibfnamefont {L.}~\bibnamefont
  {Piroli}}, \bibinfo {author} {\bibfnamefont {P.}~\bibnamefont {Calabrese}}, \
  and\ \bibinfo {author} {\bibfnamefont {F.~H.~L.}\ \bibnamefont {Essler}},\
  }\href {\doibase 10.1103/PhysRevLett.116.070408} {\bibfield  {journal}
  {\bibinfo  {journal} {Phys. Rev. Lett.}\ }\textbf {\bibinfo {volume} {116}},\
  \bibinfo {pages} {070408} (\bibinfo {year} {2016}{\natexlab{a}})}\BibitemShut
  {NoStop}%
\bibitem [{\citenamefont {Piroli}\ \emph
  {et~al.}(2016{\natexlab{b}})\citenamefont {Piroli}, \citenamefont
  {Calabrese},\ and\ \citenamefont {Essler}}]{SciPostPhys.1.1.001}%
  \BibitemOpen
  \bibfield  {author} {\bibinfo {author} {\bibfnamefont {L.}~\bibnamefont
  {Piroli}}, \bibinfo {author} {\bibfnamefont {P.}~\bibnamefont {Calabrese}}, \
  and\ \bibinfo {author} {\bibfnamefont {F.~H.~L.}\ \bibnamefont {Essler}},\
  }\href {\doibase 10.21468/SciPostPhys.1.1.001} {\bibfield  {journal}
  {\bibinfo  {journal} {SciPost Phys.}\ }\textbf {\bibinfo {volume} {1}},\
  \bibinfo {pages} {001} (\bibinfo {year} {2016}{\natexlab{b}})}\BibitemShut
  {NoStop}%
\bibitem [{\citenamefont {Panfil}\ \emph {et~al.}(2013)\citenamefont {Panfil},
  \citenamefont {De~Nardis},\ and\ \citenamefont
  {Caux}}]{PhysRevLett.110.125302}%
  \BibitemOpen
  \bibfield  {author} {\bibinfo {author} {\bibfnamefont {M.}~\bibnamefont
  {Panfil}}, \bibinfo {author} {\bibfnamefont {J.}~\bibnamefont {De~Nardis}}, \
  and\ \bibinfo {author} {\bibfnamefont {J.-S.}\ \bibnamefont {Caux}},\ }\href
  {\doibase 10.1103/PhysRevLett.110.125302} {\bibfield  {journal} {\bibinfo
  {journal} {Phys. Rev. Lett.}\ }\textbf {\bibinfo {volume} {110}},\ \bibinfo
  {pages} {125302} (\bibinfo {year} {2013})}\BibitemShut {NoStop}%
\bibitem [{\citenamefont {Iyer}\ and\ \citenamefont
  {Andrei}(2012)}]{PhysRevLett.109.115304}%
  \BibitemOpen
  \bibfield  {author} {\bibinfo {author} {\bibfnamefont {D.}~\bibnamefont
  {Iyer}}\ and\ \bibinfo {author} {\bibfnamefont {N.}~\bibnamefont {Andrei}},\
  }\href {\doibase 10.1103/PhysRevLett.109.115304} {\bibfield  {journal}
  {\bibinfo  {journal} {Phys. Rev. Lett.}\ }\textbf {\bibinfo {volume} {109}},\
  \bibinfo {pages} {115304} (\bibinfo {year} {2012})}\BibitemShut {NoStop}%
\bibitem [{\citenamefont {Iyer}\ \emph {et~al.}(2013)\citenamefont {Iyer},
  \citenamefont {Guan},\ and\ \citenamefont {Andrei}}]{PhysRevA.87.053628}%
  \BibitemOpen
  \bibfield  {author} {\bibinfo {author} {\bibfnamefont {D.}~\bibnamefont
  {Iyer}}, \bibinfo {author} {\bibfnamefont {H.}~\bibnamefont {Guan}}, \ and\
  \bibinfo {author} {\bibfnamefont {N.}~\bibnamefont {Andrei}},\ }\href
  {\doibase 10.1103/PhysRevA.87.053628} {\bibfield  {journal} {\bibinfo
  {journal} {Phys. Rev. A}\ }\textbf {\bibinfo {volume} {87}},\ \bibinfo
  {pages} {053628} (\bibinfo {year} {2013})}\BibitemShut {NoStop}%
\bibitem [{\citenamefont {Zill}\ \emph {et~al.}(2018)\citenamefont {Zill},
  \citenamefont {Wright}, \citenamefont {Kheruntsyan}, \citenamefont
  {Gasenzer},\ and\ \citenamefont {Davis}}]{SciPostPhys.4.2.011}%
  \BibitemOpen
  \bibfield  {author} {\bibinfo {author} {\bibfnamefont {J.~C.}\ \bibnamefont
  {Zill}}, \bibinfo {author} {\bibfnamefont {T.~M.}\ \bibnamefont {Wright}},
  \bibinfo {author} {\bibfnamefont {K.~V.}\ \bibnamefont {Kheruntsyan}},
  \bibinfo {author} {\bibfnamefont {T.}~\bibnamefont {Gasenzer}}, \ and\
  \bibinfo {author} {\bibfnamefont {M.~J.}\ \bibnamefont {Davis}},\ }\href
  {\doibase 10.21468/SciPostPhys.4.2.011} {\bibfield  {journal} {\bibinfo
  {journal} {SciPost Phys.}\ }\textbf {\bibinfo {volume} {4}},\ \bibinfo
  {pages} {011} (\bibinfo {year} {2018})}\BibitemShut {NoStop}%
\bibitem [{\citenamefont {Huang}(2009)}]{huang2009introduction}%
  \BibitemOpen
  \bibfield  {author} {\bibinfo {author} {\bibfnamefont {K.}~\bibnamefont
  {Huang}},\ }\href@noop {} {\emph {\bibinfo {title} {Introduction to
  statistical physics}}}\ (\bibinfo  {publisher} {CRC press},\ \bibinfo {year}
  {2009})\BibitemShut {NoStop}%
\bibitem [{\citenamefont {Saint-Jalm}\ \emph {et~al.}(2019)\citenamefont
  {Saint-Jalm}, \citenamefont {Castilho}, \citenamefont {Le~Cerf},
  \citenamefont {Bakkali-Hassani}, \citenamefont {Ville}, \citenamefont
  {Nascimbene}, \citenamefont {Beugnon},\ and\ \citenamefont
  {Dalibard}}]{PhysRevX.9.021035}%
  \BibitemOpen
  \bibfield  {author} {\bibinfo {author} {\bibfnamefont {R.}~\bibnamefont
  {Saint-Jalm}}, \bibinfo {author} {\bibfnamefont {P.~C.~M.}\ \bibnamefont
  {Castilho}}, \bibinfo {author} {\bibfnamefont {E.}~\bibnamefont {Le~Cerf}},
  \bibinfo {author} {\bibfnamefont {B.}~\bibnamefont {Bakkali-Hassani}},
  \bibinfo {author} {\bibfnamefont {J.-L.}\ \bibnamefont {Ville}}, \bibinfo
  {author} {\bibfnamefont {S.}~\bibnamefont {Nascimbene}}, \bibinfo {author}
  {\bibfnamefont {J.}~\bibnamefont {Beugnon}}, \ and\ \bibinfo {author}
  {\bibfnamefont {J.}~\bibnamefont {Dalibard}},\ }\href {\doibase
  10.1103/PhysRevX.9.021035} {\bibfield  {journal} {\bibinfo  {journal} {Phys.
  Rev. X}\ }\textbf {\bibinfo {volume} {9}},\ \bibinfo {pages} {021035}
  (\bibinfo {year} {2019})}\BibitemShut {NoStop}%
\bibitem [{\citenamefont {Kwon}\ \emph {et~al.}(2020)\citenamefont {Kwon},
  \citenamefont {Del~Pace}, \citenamefont {Panza}, \citenamefont {Inguscio},
  \citenamefont {Zwerger}, \citenamefont {Zaccanti}, \citenamefont {Scazza},\
  and\ \citenamefont {Roati}}]{Kwon84}%
  \BibitemOpen
  \bibfield  {author} {\bibinfo {author} {\bibfnamefont {W.~J.}\ \bibnamefont
  {Kwon}}, \bibinfo {author} {\bibfnamefont {G.}~\bibnamefont {Del~Pace}},
  \bibinfo {author} {\bibfnamefont {R.}~\bibnamefont {Panza}}, \bibinfo
  {author} {\bibfnamefont {M.}~\bibnamefont {Inguscio}}, \bibinfo {author}
  {\bibfnamefont {W.}~\bibnamefont {Zwerger}}, \bibinfo {author} {\bibfnamefont
  {M.}~\bibnamefont {Zaccanti}}, \bibinfo {author} {\bibfnamefont
  {F.}~\bibnamefont {Scazza}}, \ and\ \bibinfo {author} {\bibfnamefont
  {G.}~\bibnamefont {Roati}},\ }\href {\doibase 10.1126/science.aaz2463}
  {\bibfield  {journal} {\bibinfo  {journal} {Science}\ }\textbf {\bibinfo
  {volume} {369}},\ \bibinfo {pages} {84} (\bibinfo {year} {2020})},\ \Eprint
  {http://arxiv.org/abs/https://science.sciencemag.org/content/369/6499/84.full.pdf}
  {https://science.sciencemag.org/content/369/6499/84.full.pdf} \BibitemShut
  {NoStop}%
\bibitem [{\citenamefont {Castro-Alvaredo}\ \emph {et~al.}(2016)\citenamefont
  {Castro-Alvaredo}, \citenamefont {Doyon},\ and\ \citenamefont
  {Yoshimura}}]{PhysRevX.6.041065}%
  \BibitemOpen
  \bibfield  {author} {\bibinfo {author} {\bibfnamefont {O.~A.}\ \bibnamefont
  {Castro-Alvaredo}}, \bibinfo {author} {\bibfnamefont {B.}~\bibnamefont
  {Doyon}}, \ and\ \bibinfo {author} {\bibfnamefont {T.}~\bibnamefont
  {Yoshimura}},\ }\href {\doibase 10.1103/PhysRevX.6.041065} {\bibfield
  {journal} {\bibinfo  {journal} {Phys. Rev. X}\ }\textbf {\bibinfo {volume}
  {6}},\ \bibinfo {pages} {041065} (\bibinfo {year} {2016})}\BibitemShut
  {NoStop}%
\bibitem [{\citenamefont {Bertini}\ \emph {et~al.}(2016)\citenamefont
  {Bertini}, \citenamefont {Collura}, \citenamefont {{De Nardis}},\ and\
  \citenamefont {Fagotti}}]{Bertini16}%
  \BibitemOpen
  \bibfield  {author} {\bibinfo {author} {\bibfnamefont {B.}~\bibnamefont
  {Bertini}}, \bibinfo {author} {\bibfnamefont {M.}~\bibnamefont {Collura}},
  \bibinfo {author} {\bibfnamefont {J.}~\bibnamefont {{De Nardis}}}, \ and\
  \bibinfo {author} {\bibfnamefont {M.}~\bibnamefont {Fagotti}},\ }\href
  {\doibase 10.1103/PhysRevLett.117.207201} {\bibfield  {journal} {\bibinfo
  {journal} {Phys. Rev. Lett.}\ }\textbf {\bibinfo {volume} {117}},\ \bibinfo
  {pages} {207201} (\bibinfo {year} {2016})}\BibitemShut {NoStop}%
\bibitem [{\citenamefont {Doyon}\ \emph {et~al.}(2018)\citenamefont {Doyon},
  \citenamefont {Yoshimura},\ and\ \citenamefont {Caux}}]{doyonfleagas}%
  \BibitemOpen
  \bibfield  {author} {\bibinfo {author} {\bibfnamefont {B.}~\bibnamefont
  {Doyon}}, \bibinfo {author} {\bibfnamefont {T.}~\bibnamefont {Yoshimura}}, \
  and\ \bibinfo {author} {\bibfnamefont {J.-S.}\ \bibnamefont {Caux}},\ }\href
  {\doibase 10.1103/PhysRevLett.120.045301} {\bibfield  {journal} {\bibinfo
  {journal} {Phys. Rev. Lett.}\ }\textbf {\bibinfo {volume} {120}},\ \bibinfo
  {pages} {045301} (\bibinfo {year} {2018})}\BibitemShut {NoStop}%
\bibitem [{\citenamefont {Piroli}\ \emph {et~al.}(2017)\citenamefont {Piroli},
  \citenamefont {De~Nardis}, \citenamefont {Collura}, \citenamefont {Bertini},\
  and\ \citenamefont {Fagotti}}]{PhysRevB.96.115124}%
  \BibitemOpen
  \bibfield  {author} {\bibinfo {author} {\bibfnamefont {L.}~\bibnamefont
  {Piroli}}, \bibinfo {author} {\bibfnamefont {J.}~\bibnamefont {De~Nardis}},
  \bibinfo {author} {\bibfnamefont {M.}~\bibnamefont {Collura}}, \bibinfo
  {author} {\bibfnamefont {B.}~\bibnamefont {Bertini}}, \ and\ \bibinfo
  {author} {\bibfnamefont {M.}~\bibnamefont {Fagotti}},\ }\href {\doibase
  10.1103/PhysRevB.96.115124} {\bibfield  {journal} {\bibinfo  {journal} {Phys.
  Rev. B}\ }\textbf {\bibinfo {volume} {96}},\ \bibinfo {pages} {115124}
  (\bibinfo {year} {2017})}\BibitemShut {NoStop}%
\bibitem [{\citenamefont {Ilievski}\ and\ \citenamefont
  {De~Nardis}(2017{\natexlab{a}})}]{PhysRevB.96.081118}%
  \BibitemOpen
  \bibfield  {author} {\bibinfo {author} {\bibfnamefont {E.}~\bibnamefont
  {Ilievski}}\ and\ \bibinfo {author} {\bibfnamefont {J.}~\bibnamefont
  {De~Nardis}},\ }\href {\doibase 10.1103/PhysRevB.96.081118} {\bibfield
  {journal} {\bibinfo  {journal} {Phys. Rev. B}\ }\textbf {\bibinfo {volume}
  {96}},\ \bibinfo {pages} {081118} (\bibinfo {year}
  {2017}{\natexlab{a}})}\BibitemShut {NoStop}%
\bibitem [{\citenamefont {Collura}\ \emph {et~al.}(2018)\citenamefont
  {Collura}, \citenamefont {De~Luca},\ and\ \citenamefont
  {Viti}}]{collura2018analytic}%
  \BibitemOpen
  \bibfield  {author} {\bibinfo {author} {\bibfnamefont {M.}~\bibnamefont
  {Collura}}, \bibinfo {author} {\bibfnamefont {A.}~\bibnamefont {De~Luca}}, \
  and\ \bibinfo {author} {\bibfnamefont {J.}~\bibnamefont {Viti}},\ }\href
  {\doibase 10.1103/PhysRevB.97.081111} {\bibfield  {journal} {\bibinfo
  {journal} {Phys. Rev. B}\ }\textbf {\bibinfo {volume} {97}},\ \bibinfo
  {pages} {081111} (\bibinfo {year} {2018})}\BibitemShut {NoStop}%
\bibitem [{\citenamefont {Myers}\ \emph {et~al.}(2020)\citenamefont {Myers},
  \citenamefont {Bhaseen}, \citenamefont {Harris},\ and\ \citenamefont
  {Doyon}}]{10.21468/SciPostPhys.8.1.007}%
  \BibitemOpen
  \bibfield  {author} {\bibinfo {author} {\bibfnamefont {J.}~\bibnamefont
  {Myers}}, \bibinfo {author} {\bibfnamefont {M.~J.}\ \bibnamefont {Bhaseen}},
  \bibinfo {author} {\bibfnamefont {R.~J.}\ \bibnamefont {Harris}}, \ and\
  \bibinfo {author} {\bibfnamefont {B.}~\bibnamefont {Doyon}},\ }\href
  {\doibase 10.21468/SciPostPhys.8.1.007} {\bibfield  {journal} {\bibinfo
  {journal} {SciPost Phys.}\ }\textbf {\bibinfo {volume} {8}},\ \bibinfo
  {pages} {7} (\bibinfo {year} {2020})}\BibitemShut {NoStop}%
\bibitem [{\citenamefont {Ilievski}\ and\ \citenamefont
  {De~Nardis}(2017{\natexlab{b}})}]{Ilievski17}%
  \BibitemOpen
  \bibfield  {author} {\bibinfo {author} {\bibfnamefont {E.}~\bibnamefont
  {Ilievski}}\ and\ \bibinfo {author} {\bibfnamefont {J.}~\bibnamefont
  {De~Nardis}},\ }\href {\doibase 10.1103/PhysRevLett.119.020602} {\bibfield
  {journal} {\bibinfo  {journal} {Phys. Rev. Lett.}\ }\textbf {\bibinfo
  {volume} {119}},\ \bibinfo {pages} {020602} (\bibinfo {year}
  {2017}{\natexlab{b}})}\BibitemShut {NoStop}%
\bibitem [{\citenamefont {Doyon}\ and\ \citenamefont
  {Spohn}(2017{\natexlab{a}})}]{SciPostPhys.3.6.039}%
  \BibitemOpen
  \bibfield  {author} {\bibinfo {author} {\bibfnamefont {B.}~\bibnamefont
  {Doyon}}\ and\ \bibinfo {author} {\bibfnamefont {H.}~\bibnamefont {Spohn}},\
  }\href {\doibase 10.21468/SciPostPhys.3.6.039} {\bibfield  {journal}
  {\bibinfo  {journal} {SciPost Phys.}\ }\textbf {\bibinfo {volume} {3}},\
  \bibinfo {pages} {039} (\bibinfo {year} {2017}{\natexlab{a}})}\BibitemShut
  {NoStop}%
\bibitem [{\citenamefont {Bastianello}\ \emph {et~al.}(2018)\citenamefont
  {Bastianello}, \citenamefont {Doyon}, \citenamefont {Watts},\ and\
  \citenamefont {Yoshimura}}]{10.21468/SciPostPhys.4.6.045}%
  \BibitemOpen
  \bibfield  {author} {\bibinfo {author} {\bibfnamefont {A.}~\bibnamefont
  {Bastianello}}, \bibinfo {author} {\bibfnamefont {B.}~\bibnamefont {Doyon}},
  \bibinfo {author} {\bibfnamefont {G.}~\bibnamefont {Watts}}, \ and\ \bibinfo
  {author} {\bibfnamefont {T.}~\bibnamefont {Yoshimura}},\ }\href {\doibase
  10.21468/SciPostPhys.4.6.045} {\bibfield  {journal} {\bibinfo  {journal}
  {SciPost Phys.}\ }\textbf {\bibinfo {volume} {4}},\ \bibinfo {pages} {45}
  (\bibinfo {year} {2018})}\BibitemShut {NoStop}%
\bibitem [{\citenamefont {Doyon}\ and\ \citenamefont
  {Spohn}(2017{\natexlab{b}})}]{Doyon_2017}%
  \BibitemOpen
  \bibfield  {author} {\bibinfo {author} {\bibfnamefont {B.}~\bibnamefont
  {Doyon}}\ and\ \bibinfo {author} {\bibfnamefont {H.}~\bibnamefont {Spohn}},\
  }\href {\doibase 10.1088/1742-5468/aa7abf} {\bibfield  {journal} {\bibinfo
  {journal} {Journal of Statistical Mechanics: Theory and Experiment}\ }\textbf
  {\bibinfo {volume} {2017}},\ \bibinfo {pages} {073210} (\bibinfo {year}
  {2017}{\natexlab{b}})}\BibitemShut {NoStop}%
\bibitem [{\citenamefont {Doyon}(2019)}]{doi:10.1063/1.5096892}%
  \BibitemOpen
  \bibfield  {author} {\bibinfo {author} {\bibfnamefont {B.}~\bibnamefont
  {Doyon}},\ }\href {\doibase 10.1063/1.5096892} {\bibfield  {journal}
  {\bibinfo  {journal} {Journal of Mathematical Physics}\ }\textbf {\bibinfo
  {volume} {60}},\ \bibinfo {pages} {073302} (\bibinfo {year}
  {2019})}\BibitemShut {NoStop}%
\bibitem [{\citenamefont {Bulchandani}\ \emph {et~al.}(2019)\citenamefont
  {Bulchandani}, \citenamefont {Cao},\ and\ \citenamefont
  {Moore}}]{Bulchandani_2019}%
  \BibitemOpen
  \bibfield  {author} {\bibinfo {author} {\bibfnamefont {V.~B.}\ \bibnamefont
  {Bulchandani}}, \bibinfo {author} {\bibfnamefont {X.}~\bibnamefont {Cao}}, \
  and\ \bibinfo {author} {\bibfnamefont {J.~E.}\ \bibnamefont {Moore}},\ }\href
  {\doibase 10.1088/1751-8121/ab2cf0} {\bibfield  {journal} {\bibinfo
  {journal} {Journal of Physics A: Mathematical and Theoretical}\ }\textbf
  {\bibinfo {volume} {52}},\ \bibinfo {pages} {33LT01} (\bibinfo {year}
  {2019})}\BibitemShut {NoStop}%
\bibitem [{\citenamefont {Nozawa}\ and\ \citenamefont
  {Tsunetsugu}(2020)}]{PhysRevB.101.035121}%
  \BibitemOpen
  \bibfield  {author} {\bibinfo {author} {\bibfnamefont {Y.}~\bibnamefont
  {Nozawa}}\ and\ \bibinfo {author} {\bibfnamefont {H.}~\bibnamefont
  {Tsunetsugu}},\ }\href {\doibase 10.1103/PhysRevB.101.035121} {\bibfield
  {journal} {\bibinfo  {journal} {Phys. Rev. B}\ }\textbf {\bibinfo {volume}
  {101}},\ \bibinfo {pages} {035121} (\bibinfo {year} {2020})}\BibitemShut
  {NoStop}%
\bibitem [{\citenamefont {{De Nardis}}\ \emph {et~al.}(2018)\citenamefont {{De
  Nardis}}, \citenamefont {Bernard},\ and\ \citenamefont
  {Doyon}}]{PhysRevLett.121.160603}%
  \BibitemOpen
  \bibfield  {author} {\bibinfo {author} {\bibfnamefont {J.}~\bibnamefont {{De
  Nardis}}}, \bibinfo {author} {\bibfnamefont {D.}~\bibnamefont {Bernard}}, \
  and\ \bibinfo {author} {\bibfnamefont {B.}~\bibnamefont {Doyon}},\ }\href
  {\doibase 10.1103/PhysRevLett.121.160603} {\bibfield  {journal} {\bibinfo
  {journal} {Phys. Rev. Lett.}\ }\textbf {\bibinfo {volume} {121}},\ \bibinfo
  {pages} {160603} (\bibinfo {year} {2018})}\BibitemShut {NoStop}%
\bibitem [{\citenamefont {{De Nardis}}\ \emph {et~al.}(2019)\citenamefont {{De
  Nardis}}, \citenamefont {Bernard},\ and\ \citenamefont
  {Doyon}}]{10.21468/SciPostPhys.6.4.049}%
  \BibitemOpen
  \bibfield  {author} {\bibinfo {author} {\bibfnamefont {J.}~\bibnamefont {{De
  Nardis}}}, \bibinfo {author} {\bibfnamefont {D.}~\bibnamefont {Bernard}}, \
  and\ \bibinfo {author} {\bibfnamefont {B.}~\bibnamefont {Doyon}},\ }\href
  {\doibase 10.21468/SciPostPhys.6.4.049} {\bibfield  {journal} {\bibinfo
  {journal} {SciPost Phys.}\ }\textbf {\bibinfo {volume} {6}},\ \bibinfo
  {pages} {49} (\bibinfo {year} {2019})}\BibitemShut {NoStop}%
\bibitem [{\citenamefont {Gopalakrishnan}\ and\ \citenamefont
  {Vasseur}(2019)}]{PhysRevLett.122.127202}%
  \BibitemOpen
  \bibfield  {author} {\bibinfo {author} {\bibfnamefont {S.}~\bibnamefont
  {Gopalakrishnan}}\ and\ \bibinfo {author} {\bibfnamefont {R.}~\bibnamefont
  {Vasseur}},\ }\href {\doibase 10.1103/PhysRevLett.122.127202} {\bibfield
  {journal} {\bibinfo  {journal} {Phys. Rev. Lett.}\ }\textbf {\bibinfo
  {volume} {122}},\ \bibinfo {pages} {127202} (\bibinfo {year}
  {2019})}\BibitemShut {NoStop}%
\bibitem [{\citenamefont {Gopalakrishnan}\ \emph {et~al.}(2018)\citenamefont
  {Gopalakrishnan}, \citenamefont {Huse}, \citenamefont {Khemani},\ and\
  \citenamefont {Vasseur}}]{PhysRevB.98.220303}%
  \BibitemOpen
  \bibfield  {author} {\bibinfo {author} {\bibfnamefont {S.}~\bibnamefont
  {Gopalakrishnan}}, \bibinfo {author} {\bibfnamefont {D.~A.}\ \bibnamefont
  {Huse}}, \bibinfo {author} {\bibfnamefont {V.}~\bibnamefont {Khemani}}, \
  and\ \bibinfo {author} {\bibfnamefont {R.}~\bibnamefont {Vasseur}},\ }\href
  {\doibase 10.1103/PhysRevB.98.220303} {\bibfield  {journal} {\bibinfo
  {journal} {Phys. Rev. B}\ }\textbf {\bibinfo {volume} {98}},\ \bibinfo
  {pages} {220303} (\bibinfo {year} {2018})}\BibitemShut {NoStop}%
\bibitem [{\citenamefont {Cao}\ \emph {et~al.}(2018)\citenamefont {Cao},
  \citenamefont {Bulchandani},\ and\ \citenamefont
  {Moore}}]{PhysRevLett.120.164101}%
  \BibitemOpen
  \bibfield  {author} {\bibinfo {author} {\bibfnamefont {X.}~\bibnamefont
  {Cao}}, \bibinfo {author} {\bibfnamefont {V.~B.}\ \bibnamefont
  {Bulchandani}}, \ and\ \bibinfo {author} {\bibfnamefont {J.~E.}\ \bibnamefont
  {Moore}},\ }\href {\doibase 10.1103/PhysRevLett.120.164101} {\bibfield
  {journal} {\bibinfo  {journal} {Phys. Rev. Lett.}\ }\textbf {\bibinfo
  {volume} {120}},\ \bibinfo {pages} {164101} (\bibinfo {year}
  {2018})}\BibitemShut {NoStop}%
\bibitem [{\citenamefont {Medenjak}\ \emph {et~al.}(2020)\citenamefont
  {Medenjak}, \citenamefont {Nardis},\ and\ \citenamefont
  {Yoshimura}}]{10.21468/SciPostPhys.9.5.075}%
  \BibitemOpen
  \bibfield  {author} {\bibinfo {author} {\bibfnamefont {M.}~\bibnamefont
  {Medenjak}}, \bibinfo {author} {\bibfnamefont {J.~D.}\ \bibnamefont
  {Nardis}}, \ and\ \bibinfo {author} {\bibfnamefont {T.}~\bibnamefont
  {Yoshimura}},\ }\href {\doibase 10.21468/SciPostPhys.9.5.075} {\bibfield
  {journal} {\bibinfo  {journal} {SciPost Phys.}\ }\textbf {\bibinfo {volume}
  {9}},\ \bibinfo {pages} {75} (\bibinfo {year} {2020})}\BibitemShut {NoStop}%
\bibitem [{\citenamefont {Doyon}(2018)}]{10.21468/SciPostPhys.5.5.054}%
  \BibitemOpen
  \bibfield  {author} {\bibinfo {author} {\bibfnamefont {B.}~\bibnamefont
  {Doyon}},\ }\href {\doibase 10.21468/SciPostPhys.5.5.054} {\bibfield
  {journal} {\bibinfo  {journal} {SciPost Phys.}\ }\textbf {\bibinfo {volume}
  {5}},\ \bibinfo {pages} {54} (\bibinfo {year} {2018})}\BibitemShut {NoStop}%
\bibitem [{\citenamefont {Møller}\ \emph {et~al.}(2020)\citenamefont
  {Møller}, \citenamefont {Perfetto}, \citenamefont {Doyon},\ and\
  \citenamefont {Schmiedmayer}}]{10.21468/SciPostPhysCore.3.2.016}%
  \BibitemOpen
  \bibfield  {author} {\bibinfo {author} {\bibfnamefont {F.~S.}\ \bibnamefont
  {Møller}}, \bibinfo {author} {\bibfnamefont {G.}~\bibnamefont {Perfetto}},
  \bibinfo {author} {\bibfnamefont {B.}~\bibnamefont {Doyon}}, \ and\ \bibinfo
  {author} {\bibfnamefont {J.}~\bibnamefont {Schmiedmayer}},\ }\href {\doibase
  10.21468/SciPostPhysCore.3.2.016} {\bibfield  {journal} {\bibinfo  {journal}
  {SciPost Phys. Core}\ }\textbf {\bibinfo {volume} {3}},\ \bibinfo {pages}
  {16} (\bibinfo {year} {2020})}\BibitemShut {NoStop}%
\bibitem [{\citenamefont {Ruggiero}\ \emph {et~al.}(2020)\citenamefont
  {Ruggiero}, \citenamefont {Calabrese}, \citenamefont {Doyon},\ and\
  \citenamefont {Dubail}}]{ruggiero2019quantum}%
  \BibitemOpen
  \bibfield  {author} {\bibinfo {author} {\bibfnamefont {P.}~\bibnamefont
  {Ruggiero}}, \bibinfo {author} {\bibfnamefont {P.}~\bibnamefont {Calabrese}},
  \bibinfo {author} {\bibfnamefont {B.}~\bibnamefont {Doyon}}, \ and\ \bibinfo
  {author} {\bibfnamefont {J.}~\bibnamefont {Dubail}},\ }\href {\doibase
  10.1103/PhysRevLett.124.140603} {\bibfield  {journal} {\bibinfo  {journal}
  {Phys. Rev. Lett.}\ }\textbf {\bibinfo {volume} {124}},\ \bibinfo {pages}
  {140603} (\bibinfo {year} {2020})}\BibitemShut {NoStop}%
\bibitem [{\citenamefont {Bertini}\ \emph {et~al.}(2018)\citenamefont
  {Bertini}, \citenamefont {Fagotti}, \citenamefont {Piroli},\ and\
  \citenamefont {Calabrese}}]{Bertini2018}%
  \BibitemOpen
  \bibfield  {author} {\bibinfo {author} {\bibfnamefont {B.}~\bibnamefont
  {Bertini}}, \bibinfo {author} {\bibfnamefont {M.}~\bibnamefont {Fagotti}},
  \bibinfo {author} {\bibfnamefont {L.}~\bibnamefont {Piroli}}, \ and\ \bibinfo
  {author} {\bibfnamefont {P.}~\bibnamefont {Calabrese}},\ }\href {\doibase
  10.1088/1751-8121/aad82e} {\bibfield  {journal} {\bibinfo  {journal} {Journal
  of Physics A: Mathematical and Theoretical}\ }\textbf {\bibinfo {volume}
  {51}},\ \bibinfo {pages} {39LT01} (\bibinfo {year} {2018})}\BibitemShut
  {NoStop}%
\bibitem [{\citenamefont {Alba}\ \emph {et~al.}(2019)\citenamefont {Alba},
  \citenamefont {Bertini},\ and\ \citenamefont
  {Fagotti}}]{10.21468/SciPostPhys.7.1.005}%
  \BibitemOpen
  \bibfield  {author} {\bibinfo {author} {\bibfnamefont {V.}~\bibnamefont
  {Alba}}, \bibinfo {author} {\bibfnamefont {B.}~\bibnamefont {Bertini}}, \
  and\ \bibinfo {author} {\bibfnamefont {M.}~\bibnamefont {Fagotti}},\ }\href
  {\doibase 10.21468/SciPostPhys.7.1.005} {\bibfield  {journal} {\bibinfo
  {journal} {SciPost Phys.}\ }\textbf {\bibinfo {volume} {7}},\ \bibinfo
  {pages} {5} (\bibinfo {year} {2019})}\BibitemShut {NoStop}%
\bibitem [{\citenamefont {Alba}\ and\ \citenamefont
  {Calabrese}(2017)}]{alba2017entanglement}%
  \BibitemOpen
  \bibfield  {author} {\bibinfo {author} {\bibfnamefont {V.}~\bibnamefont
  {Alba}}\ and\ \bibinfo {author} {\bibfnamefont {P.}~\bibnamefont
  {Calabrese}},\ }\href {\doibase 10.1073/pnas.1703516114} {\bibfield
  {journal} {\bibinfo  {journal} {Proc. Natl. Acad. Sci. U.S.A.}\ ,\ \bibinfo
  {pages} {201703516}} (\bibinfo {year} {2017})}\BibitemShut {NoStop}%
\bibitem [{\citenamefont {Doyon}\ and\ \citenamefont
  {Yoshimura}(2017)}]{SciPostPhys.2.2.014}%
  \BibitemOpen
  \bibfield  {author} {\bibinfo {author} {\bibfnamefont {B.}~\bibnamefont
  {Doyon}}\ and\ \bibinfo {author} {\bibfnamefont {T.}~\bibnamefont
  {Yoshimura}},\ }\href {\doibase 10.21468/SciPostPhys.2.2.014} {\bibfield
  {journal} {\bibinfo  {journal} {SciPost Phys.}\ }\textbf {\bibinfo {volume}
  {2}},\ \bibinfo {pages} {014} (\bibinfo {year} {2017})}\BibitemShut {NoStop}%
\bibitem [{\citenamefont {Bastianello}\ \emph {et~al.}(2019)\citenamefont
  {Bastianello}, \citenamefont {Alba},\ and\ \citenamefont
  {Caux}}]{PhysRevLett.123.130602}%
  \BibitemOpen
  \bibfield  {author} {\bibinfo {author} {\bibfnamefont {A.}~\bibnamefont
  {Bastianello}}, \bibinfo {author} {\bibfnamefont {V.}~\bibnamefont {Alba}}, \
  and\ \bibinfo {author} {\bibfnamefont {J.-S.}\ \bibnamefont {Caux}},\ }\href
  {\doibase 10.1103/PhysRevLett.123.130602} {\bibfield  {journal} {\bibinfo
  {journal} {Phys. Rev. Lett.}\ }\textbf {\bibinfo {volume} {123}},\ \bibinfo
  {pages} {130602} (\bibinfo {year} {2019})}\BibitemShut {NoStop}%
\bibitem [{\citenamefont {Bastianello}\ and\ \citenamefont
  {De~Luca}(2019)}]{Bastianello_2019}%
  \BibitemOpen
  \bibfield  {author} {\bibinfo {author} {\bibfnamefont {A.}~\bibnamefont
  {Bastianello}}\ and\ \bibinfo {author} {\bibfnamefont {A.}~\bibnamefont
  {De~Luca}},\ }\href {\doibase 10.1103/PhysRevLett.122.240606} {\bibfield
  {journal} {\bibinfo  {journal} {Phys. Rev. Lett.}\ }\textbf {\bibinfo
  {volume} {122}},\ \bibinfo {pages} {240606} (\bibinfo {year}
  {2019})}\BibitemShut {NoStop}%
\bibitem [{\citenamefont {Caux}\ \emph {et~al.}(2019)\citenamefont {Caux},
  \citenamefont {Doyon}, \citenamefont {Dubail}, \citenamefont {Konik},\ and\
  \citenamefont {Yoshimura}}]{10.21468/SciPostPhys.6.6.070}%
  \BibitemOpen
  \bibfield  {author} {\bibinfo {author} {\bibfnamefont {J.-S.}\ \bibnamefont
  {Caux}}, \bibinfo {author} {\bibfnamefont {B.}~\bibnamefont {Doyon}},
  \bibinfo {author} {\bibfnamefont {J.}~\bibnamefont {Dubail}}, \bibinfo
  {author} {\bibfnamefont {R.}~\bibnamefont {Konik}}, \ and\ \bibinfo {author}
  {\bibfnamefont {T.}~\bibnamefont {Yoshimura}},\ }\href {\doibase
  10.21468/SciPostPhys.6.6.070} {\bibfield  {journal} {\bibinfo  {journal}
  {SciPost Phys.}\ }\textbf {\bibinfo {volume} {6}},\ \bibinfo {pages} {70}
  (\bibinfo {year} {2019})}\BibitemShut {NoStop}%
\bibitem [{\citenamefont {Friedman}\ \emph {et~al.}(2020)\citenamefont
  {Friedman}, \citenamefont {Gopalakrishnan},\ and\ \citenamefont
  {Vasseur}}]{friedman2019diffusive}%
  \BibitemOpen
  \bibfield  {author} {\bibinfo {author} {\bibfnamefont {A.~J.}\ \bibnamefont
  {Friedman}}, \bibinfo {author} {\bibfnamefont {S.}~\bibnamefont
  {Gopalakrishnan}}, \ and\ \bibinfo {author} {\bibfnamefont {R.}~\bibnamefont
  {Vasseur}},\ }\href {\doibase 10.1103/PhysRevB.101.180302} {\bibfield
  {journal} {\bibinfo  {journal} {Phys. Rev. B}\ }\textbf {\bibinfo {volume}
  {101}},\ \bibinfo {pages} {180302} (\bibinfo {year} {2020})}\BibitemShut
  {NoStop}%
\bibitem [{\citenamefont {Bastianello}\ \emph
  {et~al.}(2020{\natexlab{a}})\citenamefont {Bastianello}, \citenamefont
  {De~Luca}, \citenamefont {Doyon},\ and\ \citenamefont
  {De~Nardis}}]{PhysRevLett.125.240604}%
  \BibitemOpen
  \bibfield  {author} {\bibinfo {author} {\bibfnamefont {A.}~\bibnamefont
  {Bastianello}}, \bibinfo {author} {\bibfnamefont {A.}~\bibnamefont
  {De~Luca}}, \bibinfo {author} {\bibfnamefont {B.}~\bibnamefont {Doyon}}, \
  and\ \bibinfo {author} {\bibfnamefont {J.}~\bibnamefont {De~Nardis}},\ }\href
  {\doibase 10.1103/PhysRevLett.125.240604} {\bibfield  {journal} {\bibinfo
  {journal} {Phys. Rev. Lett.}\ }\textbf {\bibinfo {volume} {125}},\ \bibinfo
  {pages} {240604} (\bibinfo {year} {2020}{\natexlab{a}})}\BibitemShut
  {NoStop}%
\bibitem [{\citenamefont {Bastianello}\ \emph
  {et~al.}(2020{\natexlab{b}})\citenamefont {Bastianello}, \citenamefont
  {De~Nardis},\ and\ \citenamefont {De~Luca}}]{PhysRevB.102.161110}%
  \BibitemOpen
  \bibfield  {author} {\bibinfo {author} {\bibfnamefont {A.}~\bibnamefont
  {Bastianello}}, \bibinfo {author} {\bibfnamefont {J.}~\bibnamefont
  {De~Nardis}}, \ and\ \bibinfo {author} {\bibfnamefont {A.}~\bibnamefont
  {De~Luca}},\ }\href {\doibase 10.1103/PhysRevB.102.161110} {\bibfield
  {journal} {\bibinfo  {journal} {Phys. Rev. B}\ }\textbf {\bibinfo {volume}
  {102}},\ \bibinfo {pages} {161110} (\bibinfo {year}
  {2020}{\natexlab{b}})}\BibitemShut {NoStop}%
\bibitem [{\citenamefont {M\o{}ller}\ \emph {et~al.}(2021)\citenamefont
  {M\o{}ller}, \citenamefont {Li}, \citenamefont {Mazets}, \citenamefont
  {Stimming}, \citenamefont {Zhou}, \citenamefont {Zhu}, \citenamefont {Chen},\
  and\ \citenamefont {Schmiedmayer}}]{PhysRevLett.126.090602}%
  \BibitemOpen
  \bibfield  {author} {\bibinfo {author} {\bibfnamefont {F.}~\bibnamefont
  {M\o{}ller}}, \bibinfo {author} {\bibfnamefont {C.}~\bibnamefont {Li}},
  \bibinfo {author} {\bibfnamefont {I.}~\bibnamefont {Mazets}}, \bibinfo
  {author} {\bibfnamefont {H.-P.}\ \bibnamefont {Stimming}}, \bibinfo {author}
  {\bibfnamefont {T.}~\bibnamefont {Zhou}}, \bibinfo {author} {\bibfnamefont
  {Z.}~\bibnamefont {Zhu}}, \bibinfo {author} {\bibfnamefont {X.}~\bibnamefont
  {Chen}}, \ and\ \bibinfo {author} {\bibfnamefont {J.}~\bibnamefont
  {Schmiedmayer}},\ }\href {\doibase 10.1103/PhysRevLett.126.090602} {\bibfield
   {journal} {\bibinfo  {journal} {Phys. Rev. Lett.}\ }\textbf {\bibinfo
  {volume} {126}},\ \bibinfo {pages} {090602} (\bibinfo {year}
  {2021})}\BibitemShut {NoStop}%
\bibitem [{\citenamefont {Ilievski}\ \emph
  {et~al.}(2016{\natexlab{a}})\citenamefont {Ilievski}, \citenamefont
  {Medenjak}, \citenamefont {Prosen},\ and\ \citenamefont
  {Zadnik}}]{Ilievski_2016}%
  \BibitemOpen
  \bibfield  {author} {\bibinfo {author} {\bibfnamefont {E.}~\bibnamefont
  {Ilievski}}, \bibinfo {author} {\bibfnamefont {M.}~\bibnamefont {Medenjak}},
  \bibinfo {author} {\bibfnamefont {T.}~\bibnamefont {Prosen}}, \ and\ \bibinfo
  {author} {\bibfnamefont {L.}~\bibnamefont {Zadnik}},\ }\href {\doibase
  10.1088/1742-5468/2016/06/064008} {\bibfield  {journal} {\bibinfo  {journal}
  {Journal of Statistical Mechanics: Theory and Experiment}\ }\textbf {\bibinfo
  {volume} {2016}},\ \bibinfo {pages} {064008} (\bibinfo {year}
  {2016}{\natexlab{a}})}\BibitemShut {NoStop}%
\bibitem [{\citenamefont {Caux}\ and\ \citenamefont
  {Essler}(2013)}]{PhysRevLett.110.257203}%
  \BibitemOpen
  \bibfield  {author} {\bibinfo {author} {\bibfnamefont {J.-S.}\ \bibnamefont
  {Caux}}\ and\ \bibinfo {author} {\bibfnamefont {F.~H.~L.}\ \bibnamefont
  {Essler}},\ }\href {\doibase 10.1103/PhysRevLett.110.257203} {\bibfield
  {journal} {\bibinfo  {journal} {Phys. Rev. Lett.}\ }\textbf {\bibinfo
  {volume} {110}},\ \bibinfo {pages} {257203} (\bibinfo {year}
  {2013})}\BibitemShut {NoStop}%
\bibitem [{\citenamefont {Franchini}(2017)}]{franchini2017introduction}%
  \BibitemOpen
  \bibfield  {author} {\bibinfo {author} {\bibfnamefont {F.}~\bibnamefont
  {Franchini}},\ }\href@noop {} {\emph {\bibinfo {title} {An introduction to
  integrable techniques for one-dimensional quantum systems}}},\ Vol.\ \bibinfo
  {volume} {940}\ (\bibinfo  {publisher} {Springer},\ \bibinfo {year}
  {2017})\BibitemShut {NoStop}%
\bibitem [{\citenamefont {Caux}(2016)}]{Caux_2016}%
  \BibitemOpen
  \bibfield  {author} {\bibinfo {author} {\bibfnamefont {J.-S.}\ \bibnamefont
  {Caux}},\ }\href {\doibase 10.1088/1742-5468/2016/06/064006} {\bibfield
  {journal} {\bibinfo  {journal} {Journal of Statistical Mechanics: Theory and
  Experiment}\ }\textbf {\bibinfo {volume} {2016}},\ \bibinfo {pages} {064006}
  (\bibinfo {year} {2016})}\BibitemShut {NoStop}%
\bibitem [{\citenamefont {Ilievski}\ \emph
  {et~al.}(2016{\natexlab{b}})\citenamefont {Ilievski}, \citenamefont {Quinn},
  \citenamefont {Nardis},\ and\ \citenamefont {Brockmann}}]{Ilievski_2016_str}%
  \BibitemOpen
  \bibfield  {author} {\bibinfo {author} {\bibfnamefont {E.}~\bibnamefont
  {Ilievski}}, \bibinfo {author} {\bibfnamefont {E.}~\bibnamefont {Quinn}},
  \bibinfo {author} {\bibfnamefont {J.~D.}\ \bibnamefont {Nardis}}, \ and\
  \bibinfo {author} {\bibfnamefont {M.}~\bibnamefont {Brockmann}},\ }\href
  {\doibase 10.1088/1742-5468/2016/06/063101} {\bibfield  {journal} {\bibinfo
  {journal} {Journal of Statistical Mechanics: Theory and Experiment}\ }\textbf
  {\bibinfo {volume} {2016}},\ \bibinfo {pages} {063101} (\bibinfo {year}
  {2016}{\natexlab{b}})}\BibitemShut {NoStop}%
\bibitem [{\citenamefont {Sotiriadis}\ and\ \citenamefont
  {Calabrese}(2014)}]{Sotiriadis_2014}%
  \BibitemOpen
  \bibfield  {author} {\bibinfo {author} {\bibfnamefont {S.}~\bibnamefont
  {Sotiriadis}}\ and\ \bibinfo {author} {\bibfnamefont {P.}~\bibnamefont
  {Calabrese}},\ }\href {\doibase 10.1088/1742-5468/2014/07/p07024} {\bibfield
  {journal} {\bibinfo  {journal} {Journal of Statistical Mechanics: Theory and
  Experiment}\ }\textbf {\bibinfo {volume} {2014}},\ \bibinfo {pages} {P07024}
  (\bibinfo {year} {2014})}\BibitemShut {NoStop}%
\bibitem [{\citenamefont {{Ulrich Schollw\"{o}ck}}(2011)}]{Schollwoeck2011}%
  \BibitemOpen
  \bibfield  {author} {\bibinfo {author} {\bibnamefont {{Ulrich
  Schollw\"{o}ck}}},\ }\href {\doibase 10.1016/j.aop.2010.09.012} {\bibfield
  {journal} {\bibinfo  {journal} {Annals of Physics}\ }\textbf {\bibinfo
  {volume} {326}},\ \bibinfo {pages} {96} (\bibinfo {year} {2011})}\BibitemShut
  {NoStop}%
\bibitem [{\citenamefont {Clark}\ \emph {et~al.}(2015)\citenamefont {Clark},
  \citenamefont {Ha}, \citenamefont {Xu},\ and\ \citenamefont
  {Chin}}]{PhysRevLett.115.155301}%
  \BibitemOpen
  \bibfield  {author} {\bibinfo {author} {\bibfnamefont {L.~W.}\ \bibnamefont
  {Clark}}, \bibinfo {author} {\bibfnamefont {L.-C.}\ \bibnamefont {Ha}},
  \bibinfo {author} {\bibfnamefont {C.-Y.}\ \bibnamefont {Xu}}, \ and\ \bibinfo
  {author} {\bibfnamefont {C.}~\bibnamefont {Chin}},\ }\href {\doibase
  10.1103/PhysRevLett.115.155301} {\bibfield  {journal} {\bibinfo  {journal}
  {Phys. Rev. Lett.}\ }\textbf {\bibinfo {volume} {115}},\ \bibinfo {pages}
  {155301} (\bibinfo {year} {2015})}\BibitemShut {NoStop}%
\bibitem [{\citenamefont {N\"agerl}\ and\ \citenamefont
  {Landini}()}]{NagerlLandini2020pvt}%
  \BibitemOpen
  \bibfield  {author} {\bibinfo {author} {\bibfnamefont {H.}~\bibnamefont
  {N\"agerl}}\ and\ \bibinfo {author} {\bibfnamefont {M.}~\bibnamefont
  {Landini}},\ }\href@noop {} {}\bibinfo {howpublished} {Private
  communication}\BibitemShut {NoStop}%
\bibitem [{\citenamefont {Bederson}\ and\ \citenamefont
  {Walther}(2003)}]{bederson2003advances}%
  \BibitemOpen
  \bibfield  {author} {\bibinfo {author} {\bibfnamefont {B.}~\bibnamefont
  {Bederson}}\ and\ \bibinfo {author} {\bibfnamefont {H.}~\bibnamefont
  {Walther}},\ }\href@noop {} {\emph {\bibinfo {title} {Advances in atomic,
  molecular, and optical physics}}}\ (\bibinfo  {publisher} {Elsevier},\
  \bibinfo {year} {2003})\BibitemShut {NoStop}%
\bibitem [{\citenamefont {Biella}\ \emph {et~al.}(2019)\citenamefont {Biella},
  \citenamefont {Collura}, \citenamefont {Rossini}, \citenamefont {De~Luca},\
  and\ \citenamefont {Mazza}}]{Biella2019}%
  \BibitemOpen
  \bibfield  {author} {\bibinfo {author} {\bibfnamefont {A.}~\bibnamefont
  {Biella}}, \bibinfo {author} {\bibfnamefont {M.}~\bibnamefont {Collura}},
  \bibinfo {author} {\bibfnamefont {D.}~\bibnamefont {Rossini}}, \bibinfo
  {author} {\bibfnamefont {A.}~\bibnamefont {De~Luca}}, \ and\ \bibinfo
  {author} {\bibfnamefont {L.}~\bibnamefont {Mazza}},\ }\href {\doibase
  10.1038/s41467-019-12784-4} {\bibfield  {journal} {\bibinfo  {journal}
  {Nature Communications}\ }\textbf {\bibinfo {volume} {10}},\ \bibinfo {pages}
  {4820} (\bibinfo {year} {2019})}\BibitemShut {NoStop}%
\bibitem [{\citenamefont {{De Luca}}\ and\ \citenamefont
  {Mussardo}(2016)}]{De_Luca_2016}%
  \BibitemOpen
  \bibfield  {author} {\bibinfo {author} {\bibfnamefont {A.}~\bibnamefont {{De
  Luca}}}\ and\ \bibinfo {author} {\bibfnamefont {G.}~\bibnamefont
  {Mussardo}},\ }\href {\doibase 10.1088/1742-5468/2016/06/064011} {\bibfield
  {journal} {\bibinfo  {journal} {Journal of Statistical Mechanics: Theory and
  Experiment}\ }\textbf {\bibinfo {volume} {2016}},\ \bibinfo {pages} {064011}
  (\bibinfo {year} {2016})}\BibitemShut {NoStop}%
\bibitem [{\citenamefont {Vecchio}\ \emph {et~al.}(2020)\citenamefont
  {Vecchio}, \citenamefont {Bastianello}, \citenamefont {Luca},\ and\
  \citenamefont {Mussardo}}]{10.21468/SciPostPhys.9.1.002}%
  \BibitemOpen
  \bibfield  {author} {\bibinfo {author} {\bibfnamefont {G.~D. V.~D.}\
  \bibnamefont {Vecchio}}, \bibinfo {author} {\bibfnamefont {A.}~\bibnamefont
  {Bastianello}}, \bibinfo {author} {\bibfnamefont {A.~D.}\ \bibnamefont
  {Luca}}, \ and\ \bibinfo {author} {\bibfnamefont {G.}~\bibnamefont
  {Mussardo}},\ }\href {\doibase 10.21468/SciPostPhys.9.1.002} {\bibfield
  {journal} {\bibinfo  {journal} {SciPost Phys.}\ }\textbf {\bibinfo {volume}
  {9}},\ \bibinfo {pages} {2} (\bibinfo {year} {2020})}\BibitemShut {NoStop}%
\bibitem [{\citenamefont {Miao}\ \emph {et~al.}(2019)\citenamefont {Miao},
  \citenamefont {Ilievski},\ and\ \citenamefont
  {Gamayun}}]{PhysRevA.99.023605}%
  \BibitemOpen
  \bibfield  {author} {\bibinfo {author} {\bibfnamefont {Y.}~\bibnamefont
  {Miao}}, \bibinfo {author} {\bibfnamefont {E.}~\bibnamefont {Ilievski}}, \
  and\ \bibinfo {author} {\bibfnamefont {O.}~\bibnamefont {Gamayun}},\ }\href
  {\doibase 10.1103/PhysRevA.99.023605} {\bibfield  {journal} {\bibinfo
  {journal} {Phys. Rev. A}\ }\textbf {\bibinfo {volume} {99}},\ \bibinfo
  {pages} {023605} (\bibinfo {year} {2019})}\BibitemShut {NoStop}%
\end{thebibliography}%

\end{document}